\documentclass{amsart}

\usepackage{xcolor}
\definecolor{lgray}{RGB}{240,240,240}
\definecolor{webgreen}{rgb}{0,.5,0}
\definecolor{webbrown}{rgb}{.6,0,0}
\definecolor{RoyalBlue}{cmyk}{1, 0.50, 0, 0}
\usepackage[colorlinks=true, breaklinks=true, urlcolor=webbrown, linkcolor=RoyalBlue, citecolor=webgreen, backref=page]{hyperref}

\usepackage{subcaption, graphicx}
\usepackage{verbatim, setspace}
\usepackage{amsmath, amssymb, commath,bm}

\usepackage{newtxtext,newtxmath}

\newtheorem{theorem}{Theorem}[section]

\newtheorem{definition}[theorem]{Definition}
\newtheorem{remark}[theorem]{Remark}
\newtheorem*{notation}{Notation}

\newcommand{\R}		{\mathbb{R}}
\newcommand{\C}		{\mathbb{C}}
\newcommand{\N}		{\mathbb{N}}
\newcommand{\Z}		{\mathbb{Z}}

\newcommand{\re}{\mathrm{Re}}
\newcommand{\im}{\mathrm{Im}}

\renewcommand{\arg}{\mathrm{arg}}
\renewcommand{\det}{\mathrm{det}}

\newcommand{\Ai}{\mathrm{Ai}}
\newcommand{\Bi}{\mathrm{Bi}}

\newcommand{\qandq}{\quad \text{and} \quad}
\newcommand{\qorq}{\quad \text{or} \quad}

\newcommand{\RS}	{\mathfrak{S}}

\newcommand{\ii}{\mathrm{i}}
\newcommand{\dd}{\mathrm{d}}
\newcommand{\ee}{\mathrm{e}}

\begin{document}

\title[Airy solutions of P$_\mathrm{II}$ and the cubic ensemble of random matrices]{On Airy solutions of P$_\mathrm{II}$ and the complex cubic ensemble of random matrices, II}

\author{Ahmad Barhoumi}
\address{Department of Mathematics, University of Michigan, East Hall, 530~Church Street, Ann Arbor, MI 48109, USA}
\email{\href{mailto:ahmadba@kth.se}{ahmadba@kth.se}}
\curraddr{Department of Mathematics, Royal Institute of Technology (KTH), Stockholm, Sweden}

\thanks{The first author was partially supported by the European Research Council (ERC), Grant Agreement No. 101002013}

\author{Pavel  Bleher}
\address{Department of Mathematical Sciences, Indiana University Indianapolis, 402~North Blackford Street, Indianapolis, IN 46202, USA}
\email{\href{mailto:pbleher@iupui.edu}{pbleher@iupui.edu}}

\author{Alfredo Dea\~no}
\address{Department of Mathematics, Universidad Carlos III de Madrid, Avda. de la Universidad 30, 28911 Legan\'es, Madrid, Spain}
\email{\href{mailto:alfredo.deanho@uc3m.es}{alfredo.deanho@uc3m.es}}
\thanks{The  third author acknowledges financial support from Universidad Carlos III de Madrid (I Convocatoria para la Recualificaci\'on del Profesorado Universitario), from Grant PID2021-123969NB-I00, funded by MCIN/AEI/ 10.13039/501100011033, and from grant PID2021-122154NB-I00 from Spanish Agencia Estatal de Investigaci\'on.}

\author{Maxim Yattselev}
\address{Department of Mathematical Sciences, Indiana University Indianapolis, 402~North Blackford Street, Indianapolis, IN 46202, USA}
\email{\href{mailto:maxyatts@iu.edu}{maxyatts@iu.edu}}
\thanks{The research of the last author was supported in part by a grant from the Simons Foundation, CGM-706591.}





\begin{abstract}
We describe the pole-free regions of the one-parameter family of special solutions of  P$_\mathrm{II}$, the second Painlev\'e equation, constructed from the Airy functions. This is achieved by exploiting the connection between these solutions and the recurrence coefficients of orthogonal polynomials that appear in the analysis of the ensemble of random matrices corresponding to the cubic potential.
\end{abstract}

\maketitle

\section{Introduction}
This paper is a study of the so-called special function solutions of P$_\mathrm{II}$, the second Painlev\'e equation, and is a continuation of our previous work  \cite{BBDY23}. While generic solutions of Painlev\'e equations are highly transcendental, it is known that all but the first Painlev\'e equation admit special solutions which can be expressed in terms of elementary or classical special functions. The second Painlev\'e equation,
\begin{equation}
\label{P2}
q^{\prime\prime}(z) = 2q^3(z) + zq(z) + \alpha,
\end{equation}
possesses both rational solutions and solutions written in terms of the Airy functions for specific values of the parameter $\alpha\in\mathbb{C}$. In this work, we are interested in the latter, which are known to be meromorphic functions of $z\in\mathbb{C}$, with only simple poles. Our goal is to identify the pole-free regions for these solutions, by connecting these Airy solutions to the recurrence coefficients for orthogonal polynomials that arise in the complex cubic random matrix model, where  \( z \), properly rescaled, appears as a parameter.

It is heuristically well understood in the field of non-Hermitian orthogonal polynomials, but requires rigorous technical asymptotic analysis that we shall carry out in subsequent publications, that the pole-free regions of the recurrence coefficients as functions of the parameter correspond to the regions of the parameter plane where the attracting set of zeros of the orthogonal polynomials (as their degree tends to $\infty$) consists of a single Jordan arc. This is usually known in the literature as the \emph{one--cut case}. The main goal of this work is to describe the phase diagram geometrically, that is, the partition of the parameter $z$ space into different regions where the geometry of the zero-attracting set is qualitatively different, thus identifying the pole-free regions for the recurrence coefficients and for the respective Airy solutions of P$_\mathrm{II}$. 

This paper is divided into three sections. In Section~\ref{sec:airy-sols}, we give a succinct but complete construction of the Airy solutions and various related functions. Section~\ref{sec:cubic-model} is devoted to a brief description of the cubic random matrix model and the related orthogonal polynomials. There, we also state the connection between orthogonal polynomials and the Airy solutions established earlier in \cite{BBDY23}. Finally, in Section~\ref{sec:symmetric-contours}, we prove the main result of this work describing the phase diagram for the complex cubic random matrix model. This geometric analysis is strongly connected with potential theory in the complex plane, and it is an essential step (the construction of the so-called $g$ function)  in the method of nonlinear steepest descent for asymptotics of orthogonal polynomials. This full asymptotic study will eventually allow us to prove rigorously the structure of the pole free regions for Airy solutions of P$_{\rm II}$.

\section{Airy Solutions of \texorpdfstring{P$_\mathrm{II}$}{Painlev\'e-II'}}
\label{sec:airy-sols}

\begin{figure}[t]
	\centering
	\begin{subfigure}[b]{0.45\textwidth}
		\centering
		\raisebox{24 pt}{\includegraphics[width=\textwidth]{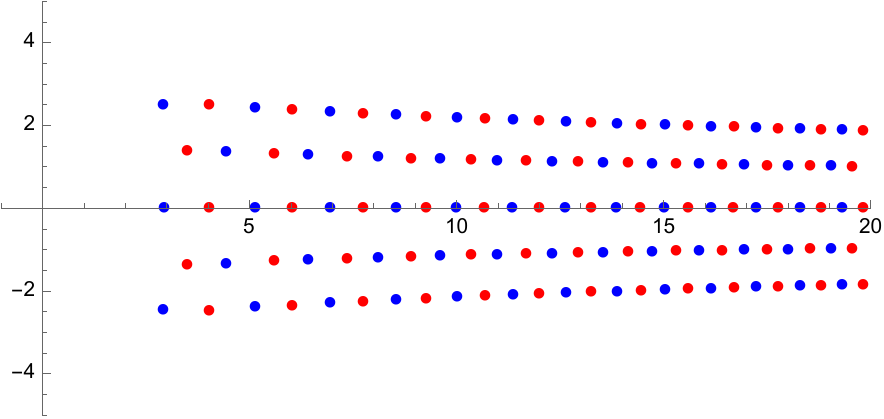}}
	\end{subfigure}
	\hfill
	\begin{subfigure}[b]{0.35\textwidth}
		\centering
		\includegraphics[width=\textwidth]{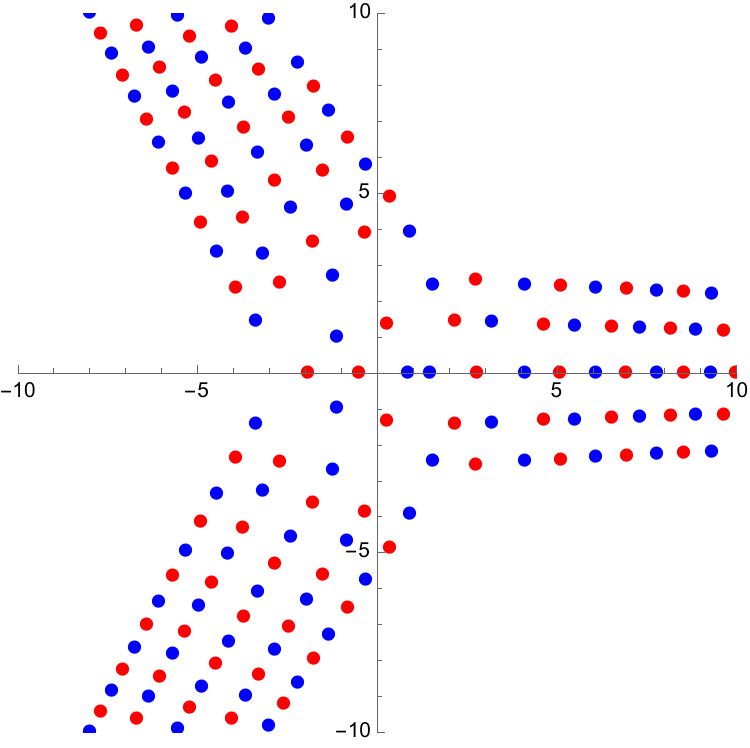}
	\end{subfigure}
	\caption{\small Zeros (blue) and poles (red) of \( q_3(z;0) \) (left panel) and \( q_3(z;\infty) \) (right panel), see \eqref{eq:q-1}--\eqref{eq:backlund-forward} for the meaning of the second parameter.}
	\label{fig:zp}
\end{figure}

It was shown by Gambier \cite{MR1555055} that P$_\mathrm{II}$ has a one-parameter family of solutions that are expressible in terms of the Airy functions and their derivatives whenever $\alpha + 1/2 = n \in \Z$. These are meromorphic functions in the complex plane whose poles exhibit a well-structured behavior, see Figure~\ref{fig:zp}.
The goal of this work is to provide a qualitative description of the pole-free regions for these solutions. Our approach proceeds through the connection with the orthogonal polynomials. Arguably one of the first successful implementations of a recasting of poles of Painelv\'e solutions as zeros of a related Hankel determinants was carried out in \cite{MR3431594}. To understand this connection, we need a determinantal representation of these solutions, see \cite{MR854008} as well as \cite{MR3529955, MR588248}, whose derivation is explained in this section.

\subsection{Airy Solutions}

 The family of the Airy solutions of P$_\mathrm{II}$ is highly structured and can be generated from a single seed solution. To obtain it, compatibility with a Riccati equation is required, i.e., it is asked that $q(z)$ also satisfies
\begin{equation}
q'(z) = k_1(z) q^2(z) + k_2(z) q(z) + k_3(z)
\label{eq:Riccati-generic}
\end{equation}
for some functions \( k_1(z),k_2(z),k_3(z) \) to be determined. Differentiating both sides of \eqref{eq:Riccati-generic} and using \eqref{P2} reduces the Riccati equation to an algebraic equation
\begin{multline*}
(2 - 2k_1^2(z))q^3(z) - (k_1'(z) + 3k_1(z)k_2(z)) q^2(z) - \\ (k_2'(z) + k_2^2(z) + 2k_1(z) k_3(z) - z) q(z) + \alpha -k_2(z)k_3(z) - k_3'(z) = 0,
\end{multline*}
which must hold identically. Setting the coefficients next to the powers of \( q(z) \) to zero gives that \( k_1(z)=\epsilon \), \( k_2(z)=0 \), \( k_3(z)=\alpha z \) while \( \alpha=\epsilon/2 \), where $\epsilon^2 = 1$. This turns \eqref{eq:Riccati-generic} into
\begin{equation}
\label{eq:Riccati-airy}
q_\epsilon^\prime(z) = \epsilon \big(q_\epsilon^2(z) + z/2\big).
\end{equation}
Choosing \(\epsilon=1 \) and making a substitution \( q_1(z) = -(\log\tau_1(z))^\prime \), linearizes \eqref{eq:Riccati-airy} and yields
\[
\tau_1''(z) + \frac 12 z \tau_1(z) = 0,
\]
which is a rescaled Airy equation that is solved by linear combinations
\begin{equation}
\label{tau1}
\tau_1(z) = C_1 \Ai(-2^{-1/3} z) + C_2 \Bi(-2^{-1/3} z),
\end{equation}
where \( C_1,C_2 \in\C \) are arbitrary constants. Thus, the corresponding solution of \eqref{eq:Riccati-airy} for \( \epsilon=1 \), i.e. \( \alpha=1/2 \), is given by
\begin{equation}
q_1(z; \lambda) = - \dod{}{z} \log \left( C_1 \, \Ai\left(-2^{-1/3}z\right) + C_2 \, \Bi\left(-2^{-1/3}z\right) \right),
\label{eq:q-1}
\end{equation}
where \( \lambda=C_2/C_1 \in \overline \C \) (note that the presence of the logarithmic derivative in the definition of $q_1(z;\lambda)$ makes it so that this is truly a one-parameter family parametrized by \( \lambda \)).  The reader might protest that imposing the Riccati condition is unmotivated, but this does have an interpretation, see Remark~\ref{remark:stationary-solution} further below.

The solutions \( q_n(z;\lambda) \) of \eqref{P2} corresponding to \( \alpha=n-1/2 \), \( n\geq 2 \), are now constructed recursively via the  B\"acklund transformation
\begin{equation}
q_{n+1}(z;\lambda) = -q_n(z;\lambda) - \frac{2n}{2q_n^2(z;\lambda)+2q_n^\prime(z;\lambda)+z},
\label{eq:backlund-forward}
\end{equation}  
see for instance \cite[32.7.25]{DLMF}. The solutions for non-positive values of \( n \) are then obtained via the symmetry \( q_n(z;\lambda) = -q_{-n+1}(z;\lambda) \).

\subsection{Hamiltonian Structure} 

Each of the first six Painlev\'e equations can be written as a Hamiltonian system
\[
\dod qz = \frac{\partial H}{\partial p} \qandq \dod pz = -\frac{\partial H}{\partial q},
\]
see \cite{MR625446,MR581468,MR596006}. In the case of P$_\mathrm{II}$ the Hamiltonian \( H_\mathrm{II} \) is equal to
\begin{equation}
\label{ham}
H_\mathrm{II}(q,p,z;\alpha) = \frac12p^2 - \left(q^2 + \frac z2\right) p - \left(\alpha+\frac12\right)q.
\end{equation}
Hence, the Hamiltonian system becomes
\begin{equation}
\label{ham_sys}
q^\prime(z) = p(z)-q^2(z) - z/2 \qandq p^\prime(z) = 2p(z)q(z) + \alpha + 1/2.
\end{equation}
Eliminating \( p(z) \) from \eqref{ham_sys} gives \eqref{P2}, while eliminating \( q(z) \) from \eqref{ham_sys} yields P$_\mathrm{XXXIV}$:
\begin{equation}
\label{P34}
p^{\prime\prime}(z) = \frac{( p^\prime(z) )^2 - (\alpha+1/2)^2}{2p(z)} + 2p^2(z) - zp(z).
\end{equation}
 Moreover, if \( q_{\alpha+1/2}(z) \) and \( p_{\alpha+1/2}(z) \) solve \eqref{P2} and \eqref{P34}, respectively, that is, as a pair solve \eqref{ham_sys}, and
\begin{equation}
\label{sigma1}
\sigma_{\alpha+1/2}(z) := H_\mathrm{II} \big( q_{\alpha+1/2}(z),p_{\alpha+1/2}(z),z;\alpha \big),
\end{equation}
then this function solves S$_\mathrm{II}$, the Jimbo-Miwa-Okamoto \( \sigma \)-form of P$_\mathrm{II}$:
\begin{equation}
\label{S2}
( \sigma^{\prime\prime})^2 + 4( \sigma^\prime )^3 + 2 \sigma^\prime ( z \sigma^\prime -\sigma ) = (\alpha/2+1/4)^2.
\end{equation}
Conversely, if \( \sigma_{\alpha+1/2}(z) \) solves \eqref{S2}, then the functions
\begin{equation}
\label{sigma2}
q_{\alpha+1/2}(z) = \frac{2\sigma_{\alpha+1/2}^{\prime\prime}(z)+\alpha+1/2}{4\sigma_{\alpha+1/2}^\prime(z)} \qandq p_{\alpha+1/2}(z) = -2\sigma_{\alpha+1/2}^\prime(z), 
\end{equation}
solve \eqref{P2} and \eqref{P34}, respectively, see \cite{MR625446,MR581468,MR596006,MR854008}. 

\begin{remark} 
\label{remark:stationary-solution}
Imposing the Riccati equation \eqref{eq:Riccati-airy} when $\epsilon = -1$, i.e., \( \alpha=-1/2 \), is equivalent to seeking a ``stationary'' solution to the Hamiltonian system where $p(z) \equiv 0$. To arrive at a similar interpretation for the choice $\epsilon = 1$ requires the following modification of the Hamiltonian, cf.\footnote{In the cited Remark, there is a sign error in the definition of $\hat H_{\mathrm{II}}$ which we correct here.} \cite[Remark 1.2]{MR854008}. Consider the change of variables
\[
\hat{q}(z) := q(z), \quad \hat p(z) := p(z) - 2q^2(z) - z,
\]
along with the Hamiltonian 
\[
\hat H_{\mathrm{II}}(\hat p, \hat q, z; \alpha) := H_{\mathrm{II}}(q, p, z; \alpha) + q = \frac12 \hat p^2 + \left( \hat q^2 + \frac z2\right) \hat p - \left(\alpha-\frac12\right) \hat q.
\]
Then, the Hamiltonian system reads 
\[
\hat q'(z) = \hat p(z) + \hat q^2(z) + z/2 \qandq \hat p'(z) = -2\hat p(z) \hat q(z) + \alpha - 1/2.
\]
Eliminating $\hat p(z)$ again gives \eqref{P2}, and now we can see that imposing the Riccati equation \eqref{eq:Riccati-airy} with $\epsilon = 1$ is equivalent to requiring $\hat p(z) \equiv 0$.
\end{remark}

\subsection{Tau Functions}

Let now \( q_n(z;\lambda) \) be the solutions of \eqref{P2} given by \eqref{eq:q-1}--\eqref{eq:backlund-forward}, while \( p_n(z;\lambda) \) and \( \sigma_n(z;\lambda) \) be the corresponding solutions of \eqref{P34} and \eqref{S2} obtained via \eqref{ham_sys} and \eqref{sigma1}, respectively.  We may now \emph{define} $\tau_n(z; \lambda)$, up to a multiplicative constant, via the formula 
\begin{equation}
\sigma_n(z; \lambda) = \dod{}{z}\log \tau_n(z; \lambda),
\label{eq:sigma-tau-n}
\end{equation}
where we take \( \tau_0\equiv 1 \) (as \( p_0\equiv 1 \) due to Remark~\ref{remark:stationary-solution},  \( \sigma_0\equiv 0 \) by \eqref{ham} and \eqref{sigma1}) and \( \tau_1(z;\lambda) \) to be \( \tau_1(z) \) from \eqref{tau1} for some choice \( C_1,C_2 \) such that \( C_2/C_1=\lambda \) (all such choices lead to the same function \( \sigma_1(z;\lambda) \)). Then
\begin{equation}
q_n(z;\lambda) \displaystyle = \dod{}{z}\log\frac{\tau_{n-1}(z;\lambda)}{\tau_n(z;\lambda)}.
\label{eq:q-tau}
\end{equation}
To arrive at expression \eqref{eq:q-tau}, we need the inverse of the B\"acklund transformation \eqref{eq:backlund-forward}, see e.g. \cite[Theorem 2]{MR3529955},
\begin{equation}
	q_{n-1}(z) = -q_n(z) + \dfrac{2(n - 1)}{2q_n'(z) - 2q_n^2(z) - z},
	\label{eq:backlund-inverse}
\end{equation}
where we drop the dependence on \( \lambda \) for brevity. Combining \eqref{eq:backlund-forward}, \eqref{eq:backlund-inverse}, and the first equation in \eqref{ham_sys}, one discovers the identity 
\begin{equation}
	p_{n - 1}(z) = -p_n(z) + 2q_n^2(z) + z.
	\label{eq:p-identity}
\end{equation}
Using \eqref{eq:p-identity}, \eqref{eq:backlund-inverse} yields the remarkable relation (cf. \cite[Eq. (1.13)\textsubscript{3}]{MR854008})
\begin{equation}
	H_{\mathrm{II}}(q_{n-1}(z), p_{n-1}(z), z; n-3/2) = H_{\mathrm{II}}(q_n(z), p_n(z), z; n-3/2),
\end{equation}
which, in view of \eqref{sigma1} and \eqref{eq:sigma-tau-n} implies \eqref{eq:q-tau}.

To derive the determinantal representation for the functions  $\tau_n(z)$, observe that they satisfy the Toda equation 
\begin{equation}
\dod[2]{}{z} \log \tau_n(z) = K_n \dfrac{\tau_{n-1}(z) \tau_{n+1}(z)}{\tau_n^2(z)}, \quad K_n \in \C.
\label{eq:toda}
\end{equation}
Indeed, on the one hand, \eqref{eq:q-tau} implies that
\begin{equation}
	q_{n+1}(z) - q_n(z) = -\dod{}{z} \log \dfrac{\tau_{n + 1}(z) \tau_{n - 1}(z)}{\tau_n^2(z)}.
	\label{eq:q-n-difference-1}
\end{equation}
On the other hand, the B\"acklund transformation \eqref{eq:backlund-forward} together with the first equation in \eqref{ham_sys} as well as \eqref{sigma2} give that
\begin{equation}
	\label{eq:q-n-difference-2}
q_{n + 1}(z) - q_n(z) = -\dfrac{n}{p_n(z)} + 2q_n(z) = -\dod{}{z} \log \sigma_n'(z).
\end{equation}
Combining \eqref{eq:q-n-difference-1}--\eqref{eq:q-n-difference-2} with \eqref{eq:sigma-tau-n} clearly implies \eqref{eq:toda}. Recall now that the tau functions are defined by \eqref{eq:sigma-tau-n} only up to a multiplicative constant. Given the seed functions \( \tau_0\) and \( \tau_1 \) as described after \eqref{eq:sigma-tau-n}, we normalize the rest of them so that \eqref{eq:toda} is satisfied with \( K_n=1 \). This yields the representation
\begin{equation}
	\label{eq:tau-n}
	\tau_n(z;\lambda) = \det \left[ \dod[j+k]{}{z} \left( C_1 \, \Ai\left(-2^{-1/3}z\right) + C_2 \, \Bi\left(-2^{-1/3}z\right) \right) \right]_{j,k=0}^{n-1},
\end{equation}
where \( C_1,C_2 \) are the same as in \eqref{tau1}, because the right-hand sides above satisfy \eqref{eq:toda} with $K_n = 1$ for all $n \geq 1$ according to Dodgson condensation identity\footnote{Also known as Desnanot-Jacobi identity or Sylvester determinant identity.}.

The functions $\tau_n(z;\lambda)$, which are entire, have an interpretation as the isomonodromic tau function, see \cite{JMU81}, that, in the context of Painlev\'e equations, were studied in great detail in \cite{MR625446}. There, the authors made the  following observation.
 
\begin{remark}
\label{remark:simplicity}
The functions $\tau_n(z;\lambda)$ have only simple zeros. Indeed, if $z_0$ is a zero of $\tau_n(z;\lambda)$ of order $p$, then, in view of \eqref{eq:sigma-tau-n}, $\sigma_n(z;\lambda)$ is a meromorphic function in $\C$ and has a simple pole at $z = z_0$ with residue $p$. A local analysis of \eqref{S2} as $z \to z_0$ reveals that 
\[
\sigma_n(z;\lambda) = \frac{1}{z - z_0} + s_0 - \dfrac{z_0}{6} (z - z_0) - \frac{1}{8}(z - z_0)^2 + \cdots, \quad s_0 \in \C,
\]
which implies that $p=1$ as claimed. 
\end{remark}

\begin{remark}
\label{remark:tau-zeros}
The functions $\tau_n(z;\lambda) \) and \( \tau_{n + 1}(z;\lambda)$ have no zeros in common for any \( n\geq 0 \). Indeed, this follows from Remark~\ref{remark:simplicity} and the fact that Toda equation \eqref{eq:toda} can be rewritten in the form	
\begin{equation}
	\tau_n(z;\lambda) \tau_n''(z;\lambda) - (\tau_n'(z;\lambda))^2 = \tau_{n + 1}(z;\lambda) \tau_{n-1}(z;\lambda).
	\label{eq:toda-2}
\end{equation}
\end{remark}

\section{Cubic Random Matrix Model}
\label{sec:cubic-model}

Cubic random matrix model studies statistics of \( N\times N \)  Hermitian matrices \( M \) drawn from the ensemble given by the probability distribution
\[
\tilde Z_N^{-1}(u) \ee^{-N\mathrm{Tr}(M^2/2-uM^3)} \dd M,
\]
where \( u \) is a parameter and \( \tilde Z_N(u) \) is the normalizing constant (partition function). This model has been investigated in physical literature by Br\'ezin, Itzykson, Parisi, and Zuber \cite{BIPZ} and Bessis, Itzykson, and Zuber \cite{BIZ}. An interesting feature of the model is that its free energy possesses an asymptotic expansion in powers of $N^{-2}$ whose coefficients are functions of $u$ which themselves admit a Taylor series expansion near $u = 0$. It was observed that the Taylor series of the coefficient of $N^{-2g}$ is the generating function for the number of three-valent graphs on a Riemann surface of genus \( g \), and so this asymptotic expansion of the free energy came to be known as the \emph{topological expansion}. The existence of a topological expansion was shown in \cite{MR1953782} for matrix models with even polynomial potential. In the case of the cubic potential this expansion was obtained in \cite{MR3071662} by the middle two authors of the present work, where they computed the number of three-valent graphs explicitly for $g = 0, 1$ and asymptotically for $g>1$.

\subsection{Connection to Airy Solutions} 
\label{subsec:D-tau-proof}

After a change of variables, see \cite{MR3071662}, the cubic ensemble turns into the unitary ensemble of random matrices whose partition function is formally defined by the matrix integral over the space of $N\times N$ Hermitian matrices:
\[
\int_{\mathcal H_N} \ee^{-N\mathrm{Tr} \left(-\frac{1}{3}M^3+tM\right)} \dd M,
\]
where \( t\in \C \) is a complex parameter. The formal partition function of the eigenvalues of these matrices can be written as
\[
\int_{-\infty}^\infty\ldots\int_{-\infty}^\infty \prod_{1\leq j<k\leq N}(s_j-s_k)^2\, \prod_{j=1}^N \ee^{-N V(s_j;t)} \dd s_1\ldots \dd s_N,
\]
with the polynomial potential \( V(s;t) \) given by
\begin{equation}
	\label{eq:V-def}
	V(s; t) := - \frac13 s^3+ st.
\end{equation}

\begin{figure}[ht!]
	\centering
	\includegraphics[scale=0.8]{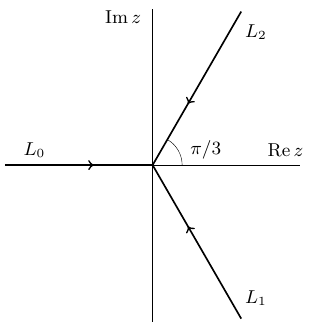}
	\caption{\small Contours $L_0$, $L_1$ and $L_2$ used in the construction of $\Gamma$.}
	\label{fig:L0L1L2}
\end{figure}
This expression is formal because the integrals are divergent and need regularization. To achieve it, let
\begin{equation}
\Gamma = \Gamma(\lambda) := \alpha_0 L_0 + \alpha_1 L_1 + \alpha_2 L_2,
\label{eq:Gamma}
\end{equation}
where \( L_k := \{z:~z=re^{(-1+2 k/3)\pi\ii},~r\in(0,\infty)\} \) are oriented towards the origin, see Figure \ref{fig:L0L1L2}, and \( \alpha_0,\alpha_1,\alpha_2\in\C \) are a one-parameter family of complex numbers given in terms of the parameter $\lambda$ (see \eqref{eq:q-1}) by
\begin{equation}
\label{alphas}
\alpha_0 = \frac\lambda\pi, \quad \alpha_1 = -\frac{\lambda}{2\pi} + \frac1{2\pi\ii}, \qandq \alpha_2 = -\frac{\lambda}{2\pi} -  \frac1{2\pi\ii},
\end{equation}
when \( |\lambda|<\infty \) and \( \alpha_0=1/\pi \), \( \alpha_1=\alpha_2 = -1/2\pi \) when \( \lambda=\infty \). 
Let
\[
Z_N(t;\lambda) := \int_{\Gamma}\ldots\int_{\Gamma} \prod_{1\leq j<k\leq N}(s_j-s_k)^2\, \prod_{j=1}^N \ee^{-N V(s_j;t)}  \dd s_1\ldots \dd s_N,
\]
which is well defined (the integrals are convergent) for all values $t\in \C$.  As briefly discussed above, it was shown in \cite{MR3071662} that the topological expansion of the free energy
\[
F_N(t;\lambda) = \frac1{N^2} \log Z_N(t;\lambda)
\]
is connected to the enumeration of regular graphs of degree 3 on Riemann surfaces. The partition function \( Z_N(t;\lambda) \) can also be expressed as \( N! D_{N-1}(t;N,\lambda) \), where
\[
D_n(t; N,\lambda) := \det \left[ \int_{\Gamma} s^{i + j } \ee^{-NV(s;t)} \dd s \right]_{i, j = 0}^n .
\]
The entries of the matrix defining \( D_n(-t;N,\lambda) \) can be written as
\begin{align*}
\int_{\Gamma} s^k \ee^{-NV(s;-t)} \dd s & = \frac1{N^{k+1/3}} \dod[k]{}{t} \left( \int_{\Gamma} \ee^{-V\left(s;-N^{2/3}t\right)} \dd s \right) \\
& = \frac1{N^{k+1/3}} \dod[k]{}{t} \left( \alpha_0\pi \, \Bi\left(-N^{2/3}t\right) + (\alpha_1-\alpha_2)\pi\ii \, \Ai\left(-N^{2/3}t\right)\right),
\end{align*}
where we used the classical integral representations, see \cite[9.5.4, 9.5.5]{DLMF}, 
\[
\Ai(z) = \frac1{2\pi\ii} \int_{L_1-L_2} \ee^{-V(s;z)} \dd s \qandq \Bi(z) = \frac1{2\pi} \left(\int_{L_0-L_1} + \int_{L_0-L_2}\right) \ee^{-V(s;z)} \dd s,
\]
as well as the condition \( \alpha_0+\alpha_1+\alpha_2=0 \). Setting \( z= (\sqrt 2N)^{2/3} t \), we get 
\begin{multline}
\int_{\Gamma} s^k \ee^{-NV(s;-t)} \dd s \\
= \frac{2^{k/3}}{N^{(k+1)/3}} \dod[k]{}{z} \left( \alpha_0\pi \, \Bi\left(-2^{-1/3}z\right) + (\alpha_1-\alpha_2)\pi\ii \, \Ai\left(-2^{-1/3}z\right)\right).
\label{eq:moments-airy}
\end{multline}
Equation \eqref{eq:moments-airy}, representation \eqref{eq:tau-n} of $\tau_n(z;\lambda)$, and the Hankel structure of the determinant \( D_n(-t;N,\lambda) \) now yield that
\begin{equation}
	\label{eq:D-tau}
	D_n(-t;N,\lambda) = \frac{2^{n(n+1)/3}}{N^{(n+1)(n+2)/3}} \tau_{n+1}(z;\lambda ),  \quad z= (\sqrt 2N)^{2/3} t.
\end{equation}
An immediate consequence of \eqref{eq:D-tau} is that these Hankel determinants satisfy a version of the Toda equation \eqref{eq:toda-2}, namely 
\[
	D_n''(t;N,\lambda) D_n(t;N,\lambda) - (D_n^\prime(t;N,\lambda))^2 = N^2 D_{n + 1}(t;N,\lambda) D_{n - 1}(t;N,\lambda). 
	\label{eq:D-toda}
\]
Furthermore, if we take \( n=N \),  then \eqref{eq:sigma-tau-n} can be rewritten in terms of the free energy \( F_N(t;\lambda) \) as
\[
\sigma_N(z;\lambda) = { N^2}\dod{}{z} F_N\left(-(\sqrt 2N)^{-2/3} z;\lambda\right),
\] 
which, of course, means that \( p_N(z;\lambda) \) and \( q_N(z;\lambda) \) can be expressed via \( F_N(t;\lambda) \) using \eqref{sigma2}. The above representations can be further rewritten using the orthogonal polynomials associated to the cubic ensemble, which we do next.

\subsection{Non-Hermitian Orthogonal Polynomials}
\label{sec:orthogonal-polynomials}

The idea of using orthogonal polynomials in computations relating to random matrices is classical at this point, see e.g. \cite{Mehta} and references therein. Let \( P_n(s;t,N,\lambda) \) be a non-identically zero polynomial of degree at most \( n \) such that
\begin{equation}
	\label{ortho}
	\int_{\Gamma} s^kP_n(s;t,N,\lambda) \ee^{-NV(s;t)} \dd s = 0,\quad k\in\{0,\ldots, n-1\},
\end{equation}
where $\Gamma=\Gamma(\lambda)$ is as in \eqref{eq:Gamma}--\eqref{alphas} and we often shall drop the explicit dependence on \( \lambda \) when this dependence is not essential to us. This family of polynomials for specific values of \( \lambda \) was studied by Huybrechs, Kuijlaars, and Lejon in \cite{MR4031470, MR3218792} in the context of complex valued Gaussian quadrature rules, and by the authors of the present work in \cite{MR4436195,MR3493550,MR3607591} in the context of the cubic model. 

Due to the non-Hermitian character of the relations in \eqref{ortho}, it might happen that polynomial satisfying \eqref{ortho} is non-unique. In this case we denote by $P_n(s;t,N,\lambda)$ the monic non-identically zero polynomial of the smallest degree; such a polynomial is always unique. It is a linear algebra exercise to verify that $\deg P_n(s; t, N,\lambda) = n$ if and only if \( D_{n - 1}(t;N) \neq 0 \). Hence, equality \eqref{eq:D-tau} and Remark~\ref{remark:tau-zeros} immediately yield the following observation.

\begin{remark}
\label{remark:polynomials-degeneration}
For fixed $t \in \C$, \( N>0 \), \( \lambda\in\overline\C \), and all $n \in \N$ it holds that
\[
\deg P_n(\cdot;t,N,\lambda) = n \qorq \deg P_{n+1}(\cdot;t,N,\lambda) = n+1. 
\]
Since \( P_{n+1}(s;t,N,\lambda) = P_n(s;t,N,\lambda) \) when \( \deg P_{n+1}(\cdot;t,N,\lambda) < n+1 \), this means that \( \deg P_{n+1}(\cdot;t,N,\lambda) \in \{ n,n+1\} \) for each \( n \).
\end{remark}

The determinantal representation of orthogonal polynomials (see, e.g., \cite{Szego}) yields that 
\[
	h_n(t;N) := \int_{\Gamma} P_n^2(s;t,N) \ee^{-NV(s;t)} \dd s = \frac{D_n(t;N)}{D_{n-1}(t;N)},
\]
where \( D_{-1}(t;N)\equiv 1 \). Since \( D_n(t;N) \) is an entire function of \( t \), each \( h_n(t;N) \) is meromorphic in \( \C \). Hence, given \( n \), the set of the values \( t \) for which there exists \( k\in\{0,\ldots,n \} \) such that \( h_k(t;N) =0 \) is countable with no limit points in the finite plane. Outside of this set a standard argument using \eqref{ortho} shows that
\begin{equation}
	\label{recurrence}
	sP_n(s;t,N) = P_{n+1}(s;t,N)+\beta_n(t;N) P_n(s;t,N)+\gamma_n^2(t;N) P_{n-1}(s;t,N),
\end{equation}
and by analytic continuation \eqref{recurrence} extends to those values of \( t \) for which \( n+1 \)-st and \( n \)-th polynomials  have the prescribed degrees (that is, \( D_{n-1}(t;N)D_n(t;N) \neq 0 \)), where
\begin{equation}\label{gammanDn}
	\gamma_n^2(t;N) = \frac{h_n(t;N)}{h_{n-1}(t;N)}=\frac{D_n(t;N)D_{n-2}(t;N)}{D_{n-1}(t;N)^2}.
\end{equation}
\sloppy Denote by \( p_{n,n-1}(t;N) \) the coefficient of \( P_n(s;t,N) \) multiplying \( s^{n-1} \) and set \( p_{0,-1}(t;N) := 0 \). It follows from \eqref{recurrence} that 
\[
\beta_n(t;N) = p_{n,n-1}(t;N)- p_{n+1,n}(t;N).
\]
Using the determinantal form for orthogonal polynomials again and the form of the weight function in our case, we can deduce that
\begin{equation}
\label{pnn1Dn}
p_{n,n-1}(t;N)
=
\frac{1}{N}\frac{\dd}{\dd t} \log D_{n-1}(t;N),
\end{equation}
and consequently
\begin{equation}
\label{betanDn}
\beta_n(t;N) 
=
\frac{1}{N}
\left(
\frac{\dd}{\dd t} \log D_{n-1}(t;N)
-
\frac{\dd}{\dd t} \log D_{n}(t;N)
\right).
\end{equation}
The following theorem has appeared in \cite{BBDY23}.
\begin{theorem}
	\label{thm:airy}
	Fix \( N \geq 1 \). Given \( \lambda\in\overline\C \), it holds for each \( n \in \N \) that
\[
		\begin{cases}
			q_n(z;\lambda) & \displaystyle = -(N/2)^{1/3} \, \beta_{n-1} \left(-(\sqrt 2N)^{-2/3} z;N,\lambda\right), \smallskip \\
			p_n(z;\lambda) & \displaystyle = -2(N/2)^{2/3} \, \gamma_n^2 \left(-(\sqrt 2N)^{-2/3} z;N,\lambda\right), \smallskip \\
			\sigma_n(z;\lambda) & \displaystyle = -(N/2)^{1/3} \, p_{n,n-1} \left(-(\sqrt 2N)^{-2/3} z;N,\lambda\right),
		\end{cases}
\]
where \( q_n(z;\lambda) \) is a solution of \eqref{P2} given by \eqref{eq:q-1} and \eqref{eq:backlund-forward}, \( p_n(z;\lambda) \) is the corresponding solution of \eqref{P34}, see \eqref{ham_sys}, while \(  \sigma_n(z;\lambda) \) is the corresponding solution of \eqref{S2} determined by \eqref{sigma1}.
\end{theorem}

As we have stated before, our goal is to identify the pole-free regions and study scaling limits of the functions \( q_n,p_n,\sigma_n \). Our approach goes through the connection to orthogonal polynomials stated in Theorem~\ref{thm:airy}. In particular, combining this theorem with \eqref{gammanDn}, \eqref{pnn1Dn}, \eqref{betanDn}, and Remarks \ref{remark:simplicity}, \ref{remark:tau-zeros} shows that the poles of the Painlev\'e function \( q_n \) in the \( z \)-plane correspond to zeros of  \( D_{n-1} \) and $D_{n}$ in the \( t \)-plane, i.e., to the values of \( t \) for which \( \deg P_n = n-1 \) or \(\deg P_{n+1} = n\), while the poles of \(p_n\) and \(\sigma_n\) correspond to zeros of \( D_{n-1} \).

The study of the asymptotic behavior of the polynomials \( P_n \) as $n\to\infty$ and for generic values of the parameter $\lambda$ (i.e. for generic combination of contours, see \eqref{eq:Gamma}) is a very technical topic that we leave to subsequent publications. However, the expected outcome of such a study is well understood: if the attracting set of zeros of \( P_n \) consists of a single Jordan arc, all the zeros of \( P_n \) remain bounded and so are the quantities \( \beta_{n-1},\gamma_n,p_{n,n-1} \); otherwise polynomials \( P_n \) will have a zero exhibiting a \emph{spurious} behavior leading to the polar singularities of \( \beta_{n-1},\gamma_n,p_{n,n-1} \).

In fact the above scheme has already been successfully carried out by us in \cite{BBDY23} where we interpreted the asymptotic results of \cite{MR4436195, MR3607591} as scaling limits of $q_n(z; \lambda)$ when $\lambda \in \{ 0, -\ii,\ii\}$. These cases are essentially the same as they correspond to taking the seed function in  \eqref{eq:tau-n} to be
\[
\Ai\left(-2^{-1/3}z\right), \quad \Ai\left(-2^{-1/3}z \ee^{2\pi\ii/3}\right),\quad \text{and} \quad \Ai\left(-2^{-1/3}z \ee^{-2\pi\ii/3}\right)
\]
since \( \Ai(z) \pm \ii \Bi(z) = 2 \ee^{\pm2\pi\ii/3} \Ai(z  \ee^{\mp2\pi\ii/3}) \), see \cite[Equation~(9.2.1)]{DLMF}. They are also qualitatively different from the remaining ones as can be easily seen from Figure~\ref{fig:zp}. Notice that these cases correspond to \( \Gamma \) in \eqref{eq:Gamma} for which \( \alpha_0\alpha_1\alpha_2 = 0 \) by \eqref{alphas}. Hence, in the remaining part of the paper we shall always assume that \( \alpha_0\alpha_1\alpha_2 \neq 0 \).

\section{Phase Diagram}
\label{sec:symmetric-contours}

In this section we take the first step in understanding the pole-free regions of Airy solutions of P$_\mathrm{II}$ corresponding to \( \lambda\notin\{0,-\ii,\ii\} \), or, equivalently, the behavior of the orthogonal polynomials corresponding to \( \Gamma \) for which \( \alpha_0\alpha_1\alpha_2 \neq 0 \). This step consists in describing the attracting set for the zeros of the orthogonal polynomials, or equivalently the phase diagram in the parameter space. For the rest of this section, we assume that \( \alpha_0\alpha_1\alpha_2 \neq 0 \).

Observe that due to the analyticity of the integrand in \eqref{ortho}, the chain of integration $\Gamma$ can be locally varied without changing the orthogonal polynomials. Hence, there arises a question of the identification of the contour attracting the zeros of the orthogonal polynomials. The answer to this question follows from Stahl-Gonchar-Rakhmanov theory of symmetric contours \cite{St85,St85b,St86,MR922628} developed in a compact setting whose implications to unbounded contours with polynomials external fields were worked out by Kuijlaars and Silva \cite{MR3306308}. 

\subsection{Symmetric Contours} 
\label{subsec:potential-theory}

Denote by $\mathcal T$ the collection of admissible contours defined as follows: each \( T\in\mathcal T \) is a finite connected union of \( C^1 \)-smooth Jordan arcs for which there exist \( R_T>0 \) and \( \epsilon_T \in(0,\pi/6) \) such that \( T\setminus\{|z|\leq R_T\} \) consists of three unbounded Jordan arcs, say \( T_0,T_1,T_2 \), each connecting a point on \( \{|z|=R_T\} \) to the point at infinity and satisfying \( T_k\subset S_{k,\epsilon_T} \), where
\[
S_{k,\epsilon} := \left \{ z : \ \left|\arg (z) + \pi - {2k\pi}/{3} \right| < \epsilon \right \}, \quad k \in\{ 0, 1, 2 \}.
\]
Notice that \( L_0\cup L_1\cup L_2 \), see \eqref{eq:Gamma}, belongs to $\mathcal{T}$. Let \( \mathcal M(T) \) be the space of Borel probability measures on $T\in\mathcal T$. The equilibrium energy of $T$ in the external field $\re \, V(\cdot;t)$, see \eqref{eq:V-def}, is equal to
\[
\mathcal E_V(T)=\inf_{\nu\in \mathcal M(T)} \left(\iint \log \frac{1}{|s-t|} \dd \nu(s) \dd \nu(t)+\int \re \, V(s;t) \dd \nu(s)\right).
\]
\begin{figure}[ht!]
	\centering
	\includegraphics[scale=1]{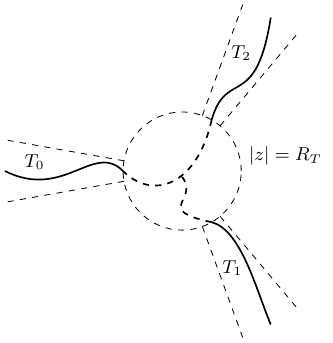}
	\caption{\small Schematic example of an admissible contour \( T\in\mathcal T \) .}
	\label{fig:curveT}
\end{figure}

The infimum above is achieved by a unique minimizer, which is called the {\it weighted equilibrium measure} of $T$ in the external field $\re \, V(\cdot;t)$, see \cite[Theorem I.1.3]{MR1485778}. We shall denote this minimizer \( \mu_T \) (of course, it is \( t \)-dependent). The support of $\mu_T$, say $J_T$, is a compact subset of $T$. The equilibrium measure  $\mu=\mu_T$ is characterized by the Euler--Lagrange variational conditions:
\begin{equation}
	\label{em2}
	2U^\mu(z)+\re \ V(z;t)\;
	\begin{cases}
		= \ell_T, & z\in J_T,\\
		\ge \ell_T, & z\in T\setminus J_T,
	\end{cases}
\end{equation}
for some constant $\ell_T$, the Lagrange multiplier, where \[ U^\mu(z):=\int\log\frac{1}{|z-s|} \dd \mu(s) \] is the logarithmic potential of $\mu$, see \cite[Theorem~I.3.3]{MR1485778}.  These notions are well defined for any \( T\in\mathcal T \). However, it is by now well understood that one should use the contour whose equilibrium measure has support  \emph{symmetric} (with the \emph{S-property}) in the external field $\re \, V(\cdot;t)$. That is, it must hold that the support $J_T$ consists of a finite number of open analytic arcs and their endpoints, and  on each arc it must hold that
\begin{equation}
	\label{em4}
	\frac{\partial }{\partial n_+}\,\big(2U^{\mu_T}+\re \, V\big)=
	\frac{\partial }{\partial n_-}\,\big(2U^{\mu_T}+\re \, V\big),
\end{equation}
where $\partial/\partial n_+$ and  $\partial/\partial n_-$ are the normal derivatives from the $(+)$- and $(-)$-side of $T$. It is said that a curve $T\in\mathcal T$ is an \emph{S-curve} in the field $\re \, V(\cdot;t)$, if $J_T$ has the S-property in this field. It is also understood that geometrically $J_T$ is comprised of \emph{critical trajectories} of a certain quadratic differential; this was already known to Stahl (see, e.g. \cite[Theorem 1]{St85b}) and remains a useful recasting of the S-property in the context of Painlev\'e transcendents, see e.g. the recent work \cite{BM22}. Recall that if $Q(z)$ is a meromorphic function, a \emph{trajectory} (resp. \emph{orthogonal trajectory}) of a quadratic differential $-Q(z) \dd z^2$ is a maximal regular arc on which
\[
-Q(z(s))\big(z^\prime(s)\big)^2>0 \quad \big(\text{resp.} \quad -Q(z(s))\big(z^\prime(s)\big)^2<0\big)
\]
for any local uniformizing parameter. A trajectory is called \emph{critical} if it is incident with a \emph{finite critical point} (a zero or a simple pole of $-Q(z) \dd z^2$) and it is called \emph{short} if it is non-recurrent and incident only with finite critical points. We designate the expression \emph{critical (orthogonal) graph of $-Q(z) \dd z^2$} for the totality of the critical (orthogonal) trajectories $-Q(z) \dd z^2$. The following theorem is a specialization to $V(z;t)$ of the results of Kuijlaars and Silva \cite[Theorems~2.3 and~2.4]{MR3306308}.

\begin{theorem}
	\label{fundamental} 
	Let $V(z;t)$ be as in \eqref{eq:V-def}. For each $t \in \C$ there exists a contour $\Gamma_t\in\mathcal T$ such that \( \mathcal E_V(\Gamma_t)=\sup_{T\in\mathcal T} \mathcal E_V(T) \).  Moreover,
	  \begin{itemize}
	  \item the support \( J_t \) of the equilibrium measure \( \mu_t :=\mu_{\Gamma_t} \) has the S-property in the external field $\re \, V(\cdot;t)$;
	  \item the function
	\begin{equation}
		\label{em5}
		Q(z;t)=\left(\frac{V'(z;t)}{2}- \int \frac{\mathrm d\mu_t(s)}{z-s}\right)^2,\quad z\in \C\setminus J_t,
	\end{equation}
	is a polynomial of degree 4;
	\item the support $J_t$ consists of some (possibly not all) short critical trajectories of the quadratic differential 
	\[
	\varpi_t(z) := -Q(z;t) \dd z^2;
	\]
	\item the equilibrium measure $\mu_t$ is absolutely continuous with respect to Lebesgue measure and
	\begin{equation}
	\dd \mu_t(z)=-\frac{1}{\pi \ii}\,Q_+^{1/2}(z;t) \dd z, \quad z\in J_t,
	\label{eq:eq-measure}
	\end{equation}
	where $Q^{1/2}(z;t)=\frac12z^2+\mathcal{O}(z)$ as $z\to\infty$ is the branch holomorphic in \( \C\setminus J_t \).
	\end{itemize}
\end{theorem}

\begin{remark}
\label{remark:basic_geo}
Recall that every \( T\in \mathcal T \) is a finite connected union of \( C^1 \)-smooth Jordan arcs. Observe also that \( \mathcal E_V(T^\prime) \geq \mathcal E_V(T) \) if \( T^\prime\subseteq T \) by the very definition of minimal energy. Hence, every extremal \( \Gamma_t \) must consist of three semi-unbounded non-intersecting open Jordan arcs  stretching out to infinity within the sectors \( S_{0,\epsilon_{\Gamma_t}},S_{1,\epsilon_{\Gamma_t}},S_{2,\epsilon_{\Gamma_t}} \) that have a common endpoint (loops, intervals of overlap, and chains of Jordan arcs with a finite terminal point can be removed without decreasing maximality of minimal energy). Thus, \( \Gamma_t = \Gamma_t^{02} \cup \Gamma_t^{12} \), where \( \Gamma_t^{ij} \) is a Jordan arc stretching to infinity within the sectors \( S_{i,\epsilon_{\Gamma_t}} \) and \( S_{j,\epsilon_{\Gamma_t}} \).  Therefore,
\[
\Gamma = \alpha_0 L_0 + \alpha_1 L_1 + \alpha_2 L_2 = \alpha_0 (L_0-L_2) + \alpha_1 (L_1-L_2)
\]
is homologous to \( \alpha_0\Gamma_t^{02} + \alpha_1 \Gamma_t^{12} \), recall \eqref{eq:Gamma}--\eqref{alphas}. Hence, we can use the latter chain in \eqref{ortho} to define orthogonal polynomials \( P_n(s;t,N,\lambda) \).
\end{remark}

\begin{remark}
\label{remark:freedom}
S-curve \( \Gamma_t \) cannot be fixed uniquely since one has freedom in choosing \( \Gamma_t \) away from \( J_t \). Indeed, let
\begin{equation}
	\label{em0}
	\mathcal U(z;t) := \re\left(2\int_e^z Q^{1/2}(s;t) \dd s \right) = \ell_{\Gamma_t} - \mathrm{Re}(V(z;t)) - 2U^{\mu_t}(z),
\end{equation}
where \( e\in J_t \) is arbitrary and the second equality follows from \eqref{em5} (since the constant \( \ell_{\Gamma_t} \) in \eqref{em2} is the same for all connected components of \( J_t \) and the integrand is purely imaginary on \( J_t \), the choice of \( e \) is indeed not important). It follows from the variational condition \eqref{em2} that \( J_t \) is a part of the set \( \{z:\mathcal U(z;t)=0 \} \) while \( \Gamma_t\setminus J_t \) must belong to \( \{z:\mathcal U(z;t)\leq 0 \} \)  and can be varied freely within this region. As \( \mathcal U(z;t) \) is harmonic in \( \C\setminus J_t \),  \( \Gamma_t\setminus J_t \) belongs to the closure of \( \{z:\mathcal U(z;t)< 0 \} \), which means that such variations are indeed possible (on the other hand, the arcs in \( J_t \) border \( \{z:\mathcal U(z;t)> 0 \} \) on both sides).
\end{remark}

\begin{remark}
\label{remark:uniqueness}
The set \( J_t \) and the measure \( \mu_t \) are unique for each parameter \( t \). Indeed, let \( \Gamma_t^* \) and \( \Gamma_t^{**} \) be two S-curves corresponding to the same \( t \) with equilibrium measures \( \mu_t^* \) and \( \mu_t^{**} \), respectively. Let \( \epsilon>0 \) be such that both \( \Gamma_t^* \) and \( \Gamma_t^{**} \) extend to infinity within \( S_{k,\epsilon} \), \( k\in\{0,1,2\} \). Choose \( R>0 \) large enough so that \( J_t^*,J_t^{**} \subset \{|z|<R\} \) while
\[
S_{k,\epsilon}\cap \{|z|\geq R\} \subset U = \big\{z:\mathcal U^*(z;t)< 0 \big\} \cap \big\{z:\mathcal U^{**}(z;t)< 0 \big \}.
\] 
According to Remark~\ref{remark:freedom}, we can modify \( \Gamma_t^* \) and \( \Gamma_t^{**} \) within \( U \) so that they intersect \( \{|z| = R \} \) at the same three points. Set \( \Gamma_{t,R}^* := \Gamma_t^* \cap \{|z|\leq R\} \) and define \( \Gamma_{t,R}^{**} \) similarly. Since \eqref{em2} characterizes weighted equilibrium measures, \( \mu_t^* \) and \( \mu_t^{**} \) remain being such measures for \( \Gamma_{t,R}^* \) and \( \Gamma_{t,R}^{**} \), respectively. As \eqref{em4} also remains valid for these measures,  \( \Gamma_{t,R}^* \) and \( \Gamma_{t,R}^{**} \) are compact S-curves. Due to Remark~\ref{remark:basic_geo} and our local modification of \( \Gamma_t^* \) and \( \Gamma_t^{**} \), \( \Gamma_{t,R}^* \) and \( \Gamma_{t,R}^{**} \) are homologous to each other and therefore non-Hermitian orthogonal polynomials on them must be the same. Then it follows from \cite[Lemma~1]{MR922628} that the normalized zero counting measures of these polynomials must simultaneously converge weak$^*$ to \( \mu_t^* \) and \( \mu_t^{**} \), which establishes the desired claim.
\end{remark}

\begin{remark}
\label{remark:symmetries}
It is enough to describe \( \mu_t \) for \( t\in O \) for any \( O \) such that \( \C=O\cup \eta O \cup \bar\eta O \), where \( \eta=\ee^{2\pi\ii/3} \). Indeed, it holds that $\eta T\in \mathcal T \) when \( T \in \mathcal T \). Let \( \mu \) be a measure on \( T \) and \( \mu^\eta \) be the push forward measure of \( \mu \) on \( \eta T \), that is, \( \mu^\eta(A) = \mu(\bar\eta A) \) for any Borel subset \( A \) of \( \eta T \). Since
\[
U^{\mu^\eta}(z) = \int \log\frac1{|z-\eta x|}d\mu(x) = U^\mu(\bar\eta z),
\]
and \( V(z;\bar \eta t)=V(\bar\eta z;t) \), it follows from \eqref{em2} that the weighted equilibrium measure of \( \eta T\) in the external field \( \re\,V(\cdot;\bar\eta t) \) is equal to \( \mu_T^\eta \), the push forward of the weighted equilibrium measure of \( T\) in the external field \( \re\,V(\cdot;t) \). This means that
\[
\sup_{T\in\mathcal T} \mathcal E_{V(\cdot;t)}(T) = \sup_{T\in\mathcal T} \mathcal E_{V(\cdot;\bar\eta t)}(T)
\]
and that if \( \Gamma_t \) is a symmetric contour from Theorem~\ref{fundamental} then \( \eta J_t = J_{\bar\eta t} \). In particular,  \( Q(z;\bar \eta t) = \eta Q(\bar\eta z;t) \).
\end{remark}

\subsection{Phase Diagram}
\label{subsec:descritpion-of-O}

Both global and local structures of the critical graph of quadratic differentials have been well-studied; see e.g.  \cite{Jenkins,Pommerenke,Strebel}. Since $\deg\, Q(z;t)=4$, $J_t$ consists of one arc, two arcs, or three arcs with a common endpoint; the first case occurs when $Q(z;t)$ has two simple and one double zero, and the other two occur when $Q(z;t)$ has four simple zeros. Below, we examine when these distinct possibilities happen.

By developing the right hand side of \eqref{em5} into a Laurent series and recalling that $Q(z;t)$ is a polynomial, we may write
\begin{equation}
Q(z; t) = \dfrac{1}{4}(z^2 - t)^2 + z + K(t),
\label{eq:Q-unfactored}
\end{equation}
for some $K(t) \in \C$. To find parameters $t$ for which $\mu_t$ is supported on a single arc (often referred to as a ``cut''), we start with the ansatz 
\begin{equation}
	Q(z;t) = \frac{1}{4}(z-a(t))(z - b(t)) (z - c(t))^2.
	\label{eq:Q-factorized-1}
\end{equation}
One could identify values of $K(t)$ for which \eqref{eq:Q-unfactored} has multiple roots by computing the discriminant, but the following approach appears more beneficial. Comparing coefficients of \eqref{eq:Q-unfactored}, \eqref{eq:Q-factorized-1} yields the system (we suppress the $t$-dependence for brevity)
\begin{equation}
\begin{cases}
		a + b + 2c &= 0, \\
		ab + c^2 + 2(a+b)c &= -2t, \\
		2abc + (a+b)c^2 &= -4.
\end{cases}
\label{eq:abc-system}
\end{equation}
Letting $x := -c = (a + b)/2$ and eliminating $ab$ from the second and third equation yields 
\begin{equation}
	x^3 - tx - 1 = 0.
	\label{eq:x-cubic}
\end{equation}
Hence, the matter of solving the system \eqref{eq:abc-system} is now translated into that of understanding the solutions of \eqref{eq:x-cubic}. Precisely, we seek to identifying values of $t\in \C$ for which the critical graph of $\varpi_t(z)$, as defined by \eqref{eq:Q-factorized-1} and \eqref{eq:abc-system}, admits a contour $\Gamma_t \in \mathcal T$ such that $\Gamma_t \subset \{ z  : \mathcal U (z;t) \leq 0 \}$ and $\mathcal U(z;t) = 0$ holds on a subset of $\Gamma_t$ of finite arclength. It was argued in \cite[Section~5]{MR3607591} that this condition can be imposed by requiring absence of a short trajectory connecting $b$ to $c$, or
\[
 \re \left(\int_b^c Q^{1/2}(s;t) \dd s\right) \neq 0.
\]
Thus, transitions in the critical graph correspond to choices of $x$ for which
\[
\re \left(\dfrac{2}{3} \int_{-1}^{2x^3} \left(1+\frac1s \right)^{3/2} \dd s \right) = 0,
\]
i.e., one needed to study the critical graph of the auxiliary quadratic differential 
\[
-\left(1+\frac1s \right)^3 \dd s^2, 
\]
denoted by $\mathcal C$.
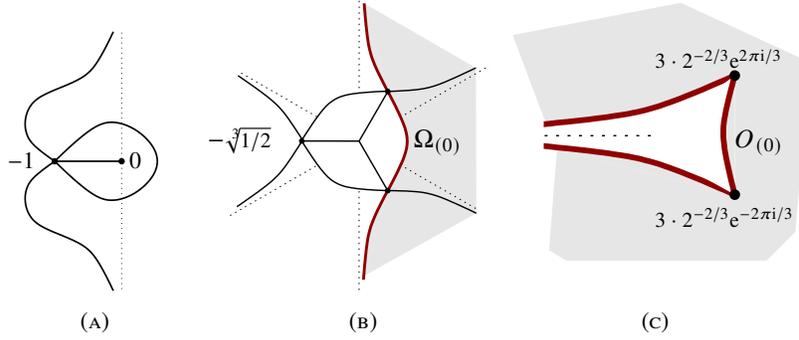
\begin{figure}[t]
	\centering
	\begin{subfigure}[b]{0.25 \textwidth}\centering
		\includegraphics[scale=.35]{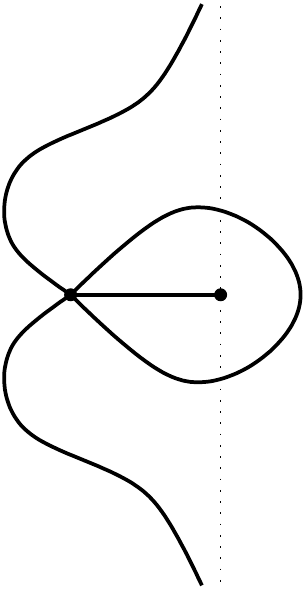}
		\begin{picture}(0,0)
			\put(-14,47){\small $0$}
			\put(-60,47){\small $-1$}
		\end{picture}
	\caption{}
	\end{subfigure}
	\begin{subfigure}[b]{0.3 \textwidth}\centering
		\includegraphics[scale=.6]{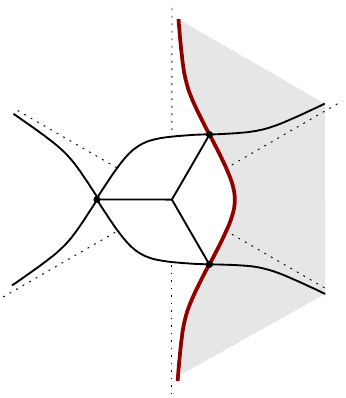}
		\begin{picture}(0,0)
			\put(-110,55){$\footnotesize -\sqrt[3]{1/2}$}
			\put(-32,55){$\Omega_{(0)}$}
		\end{picture}
	\caption{}
	\end{subfigure}
	\begin{subfigure}[b]{0.3 \textwidth}\centering
		\includegraphics[scale=1.2]{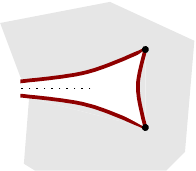}
		\begin{picture}(0,0)
			\put(0,25){\footnotesize $3\cdot2^{-2/3} \ee^{-2\pi\ii/3} $}
			\put(0,85){\footnotesize $3\cdot2^{-2/3} \ee^{2\pi\ii/3}$}
			\put(30,55){$O_{(0)}$}
		\end{picture}
	\caption{}
	\end{subfigure}
	\caption{\small Schematic representation of (A)  the critical graph $\mathcal C$; (B) the set $\Delta$ (solid lines) and the domain $\Omega_{(0)}$ (shaded region); (C) domain $O_{(0)}$ (shaded region).}
	\label{fig:loops}
\end{figure}
 It was shown in \cite{MR3607591} that $\mathcal C$ consists of 5 critical trajectories emanating from $-1$ at the angles $2\pi k/5$, $k\in\{0,1,2,3,4\}$, one of them being $(-1,0)$, two forming a loop crossing the real line approximately at $0.635$, and two approaching infinity along the imaginary axis without changing the half-plane (upper or lower), see Figure~\ref{fig:loops}(A). Define \[ \Delta:=\big\{x:~2x^3\in\mathcal C\big\} \] and let $\Omega_{(0)}$ to be the subset of the right-half plane bounded by three smooth subarcs of \( \Delta \) as on Figure~\ref{fig:loops}(B). Set \( \Omega_{(\pm \ii)} := \ee^{\mp2\pi \ii/3}\Omega_{(0)}\). The function $t(x)$ defined by \eqref{eq:x-cubic} is holomorphic in each $\Omega_{(\tau)}$, \( \tau\in\{-\ii,\ii,0\} \), with non-vanishing derivative there. Put \( O_{(\tau)} := t(\Omega_{(\tau)}) \), see Figure~\ref{fig:loops}(C), then the inverse map \( x_\tau: O_{(\tau)}\to\Omega_{(\tau)} \) exists and is holomorphic, \( \tau\in\{-\ii,\ii,0\} \), and \(  O_{(\pm \ii)} = \ee^{\pm2\pi \ii/3} O_{(0)} \).

\begin{figure}[ht!]
	\centering
	\includegraphics[scale=1]{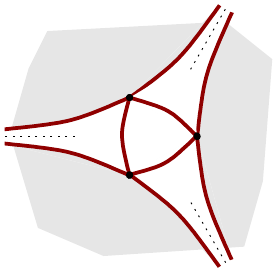}
	\begin{picture}(0,0)
		\put(-32,62){$O_{0,-\ii}$}
		\put(-95,95){$O_{0,0}$}
		\put(-90,22){$O_{0,\ii}$}
		\put(-70,62){$O_{1,-}$}
		\put(-62,83){$O_{1,-\ii}$}
		\put(-97,62){$O_{1,0}$}
		\put(-59,41){$O_{1,\ii}$}
	\end{picture}
	\caption{\small Single cut domains \( O_{0,0} \), \( O_{0,\ii} \), and \( O_{0,-\ii} \); double cut domains \( O_{1,0} \), \( O_{1,\ii} \), \( O_{1,-\ii} \); and three cuts with a common endpoint domain \( O_{1,-} \). Boundaries of these domains correspond to single cut cases.}
	\label{fig:t-plane}
\end{figure}
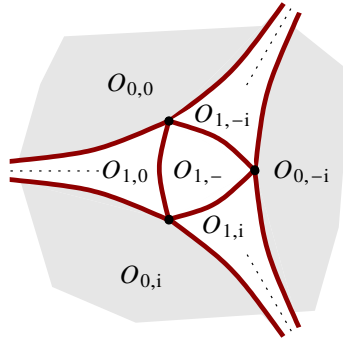

Denote by \( O_0 \) the union of the unbounded components of \( O_{(0)} \cap O_{(\ii)} \cap O_{(-\ii)} \) and by \( O_{1,-} \) the bounded component of this intersection. Further,  let \( O_{1,\tau} \) be the complement of the closure of  \( O_{(\tau)} \) and \( O_{0,\tau} \) be the connected component of \( O_0 \) that lies directly clockwise from \( O_{1,\tau} \), \( \tau\in\{0,\ii,-\ii\} \), see Figure~\ref{fig:t-plane} (numbers \( 0 \) and \( 1 \) stand for the genus of the Riemann surface of \( Q^{1/2}(z;t) \)). We further set \( O_1 := O_{1,0}\cup O_{1,\ii} \cup O_{1,-\ii} \cup O_{1,-} \). Then the following theorem takes place.

\begin{theorem}
\label{phase_diagram}
For each \( t\in\C\), let \( \Gamma_t\in\mathcal T \) be as in Theorem~\ref{fundamental}. Recall that \( J_t \), the support of the equilibrium measure \( \mu_t \), is uniquely determined. It holds that
\begin{itemize}
\item for \( t \in O_0\cup\{t_\mathsf{cr},\eta t_\mathsf{cr},\bar\eta t_\mathsf{cr}\} \) the support \( J_t \) consists of a single analytic Jordan arc, where \( \eta=\ee^{2\pi\ii/3} \) and \( t_\mathsf{cr}=3\cdot2^{-2/3} \); 
\item for \( t\in O_{1,-} \) the support \( J_t \) consists of three analytic Jordan arcs with a common endpoint;
\item for \( t \in O_1\setminus O_{1,-} \) the support \( J_t \) consists of two analytic Jordan arcs with no endpoints in common;
\item in the remaining cases the support \( J_t \) consists of two analytic Jordan arcs with a common endpoint that form a corner at it.
\end{itemize}
\end{theorem}

The claims of Theorem~\ref{phase_diagram} follows from Theorems~\ref{thm:geometry1}--\ref{thm:geometry3} further below, in which we discuss the detailed structure of these contours \( \Gamma_t \).
\begin{figure}[t]
	\includegraphics[scale = 0.5]{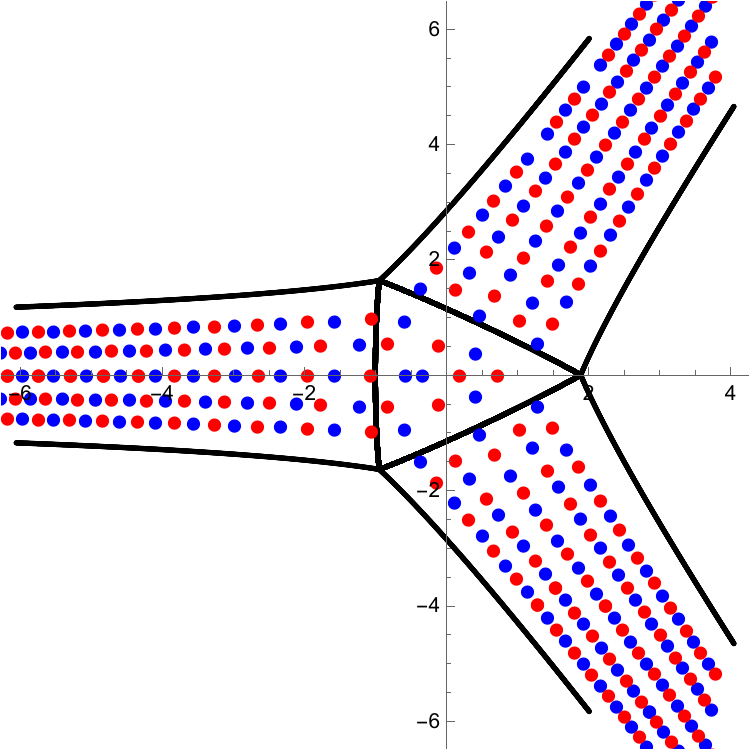}
	\caption{Poles (red) and zeros (blue) of $q_3(-(3\sqrt{2})^{2/3}z; \infty)$ and the curves $\partial O_1$ (black). }
	\label{fig:zeros-poles-contour}
\end{figure}

\begin{remark}
According to our heuristics, Airy solutions \( q_n(z;\lambda) \) must be pole-free on compact subsets of \( -(\sqrt 2n)^{2/3} O_0 \) for all \( n \) large, see Theorem{}~\ref{thm:airy} where we take \( n=N \) and Figure \ref{fig:zeros-poles-contour}. We emphasize that while the locations of the poles of $q_n(-(\sqrt 2n)^{-2/3}z; \lambda)$ depend on $\lambda \notin \{\pm \ii, 0\}$, the region which they fill in the large $n$ limit, $O_1$, does not!
\end{remark}

\subsection{One-cut Cases}
Below, we describe the structure of the critical trajectories of the quadratic differential \( \varpi_t(z) \) as well as our preferred choice of \( \Gamma_t \), where we adopt the following convention.

\begin{figure}[t]
	\begin{subfigure}[b]{0.3 \textwidth}\centering
		{\includegraphics[scale=.2]{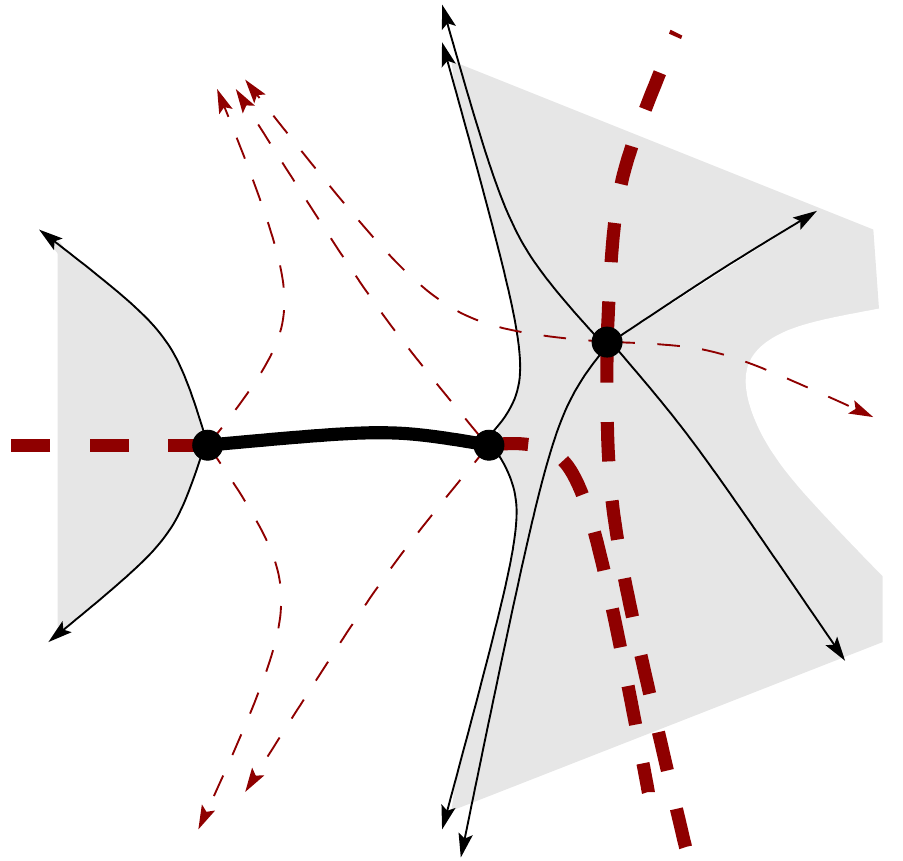}}
		\caption{}
	\end{subfigure}
	\begin{subfigure}[b]{0.3 \textwidth}\centering
		{\includegraphics[scale=.2]{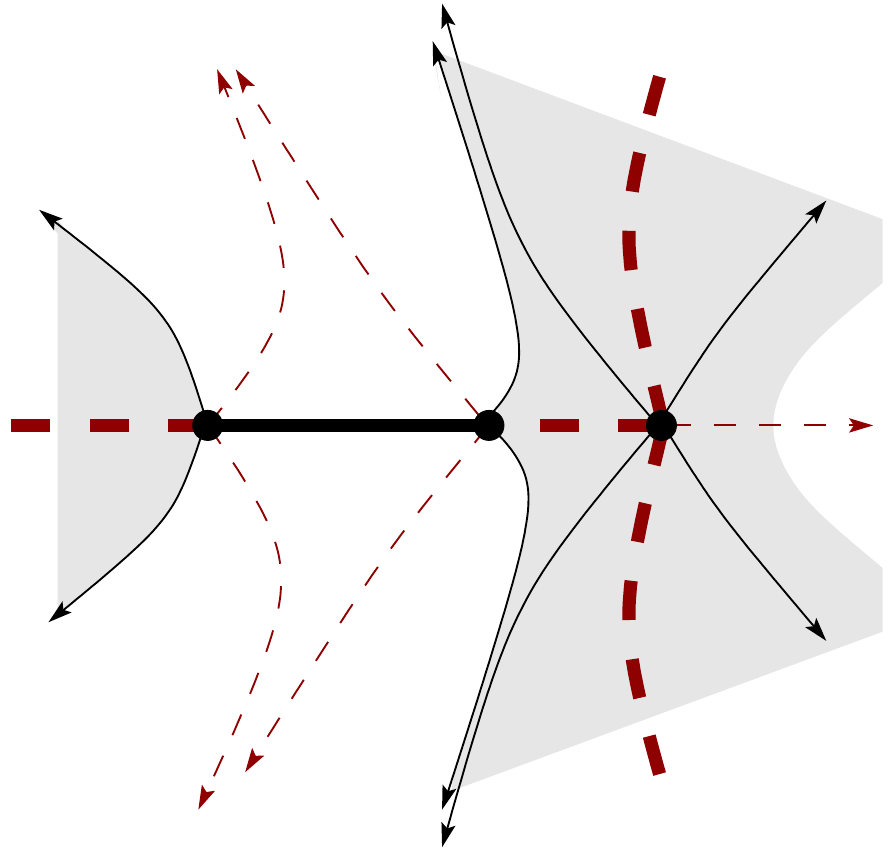}} 
		\caption{}
	\end{subfigure}
	\begin{subfigure}[b]{0.3 \textwidth}\centering
		{\includegraphics[scale=.2]{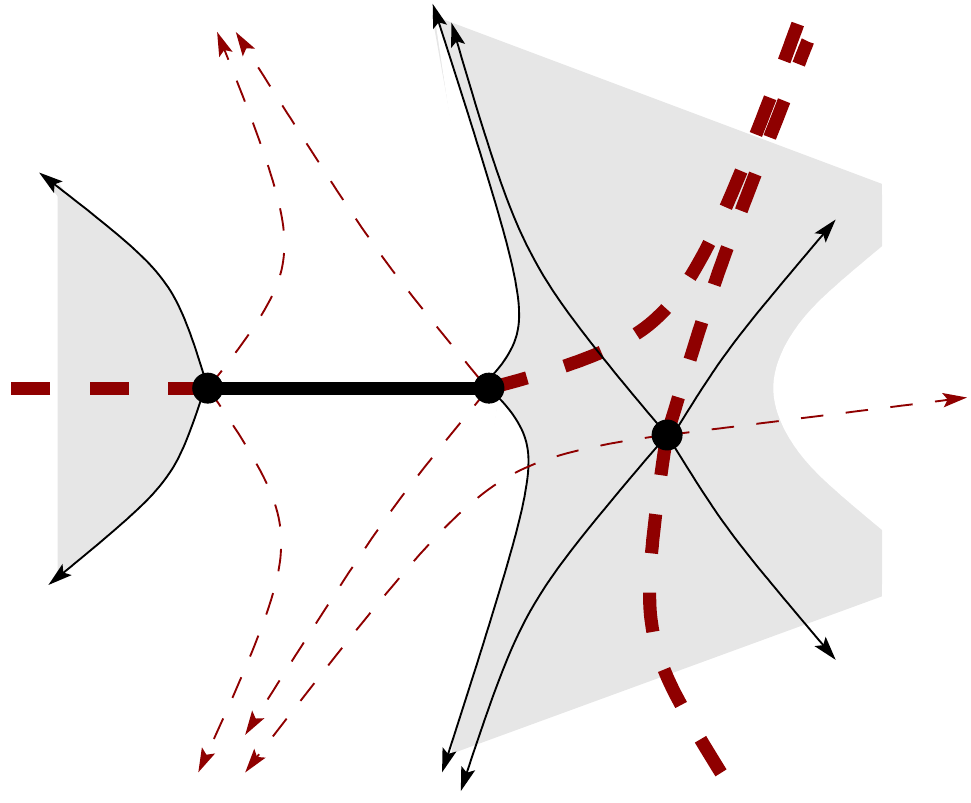}}
		\caption{}
	\end{subfigure} 
	\caption{\small Schematic representation of the critical (solid) and critical orthogonal (dashed) graphs of $\varpi_t(z)$ when (A) \( t\in  O_{0,-\ii} \), $\im(t)>0$; (B) \( t\in O_{0,-\ii} \), \( \im(t)=0 \); (C) \( t\in O_{0,-\ii} \), \( \im(t)<0 \). The bold curves represent the preferred S-curve $\Gamma_t$. Shaded region is the set $\{\mathcal U(z;t)<0\}$.}
	\label{s-curves1}
\end{figure} 

\begin{notation}\label{notation:trajectories}
$\Gamma(z_1,z_2)$ (resp. $\Gamma[z_1,z_2]$) stands for the trajectory or orthogonal trajectory (resp. the closure of) of the differential \( \varpi_t(z) \) connecting $z_1$ and $z_2$, oriented from $z_1$ to $z_2$, and $\Gamma\big(z, \ee^{\ii\theta}\infty\big)$ (resp. $\Gamma\big(\ee^{\ii\theta}\infty,z\big)$) stands for the orthogonal trajectory incident with $z$, approaching infinity at the angle $\theta$, and oriented away from $z$ (resp. oriented towards $z$).\footnote{This notation is unambiguous as the corresponding trajectories are unique for polynomial differentials as follows from Teichm\"uller's lemma.}
\end{notation}

\begin{figure}[t]
	\begin{subfigure}[b]{0.3 \textwidth}\centering
		\includegraphics[scale=.2]{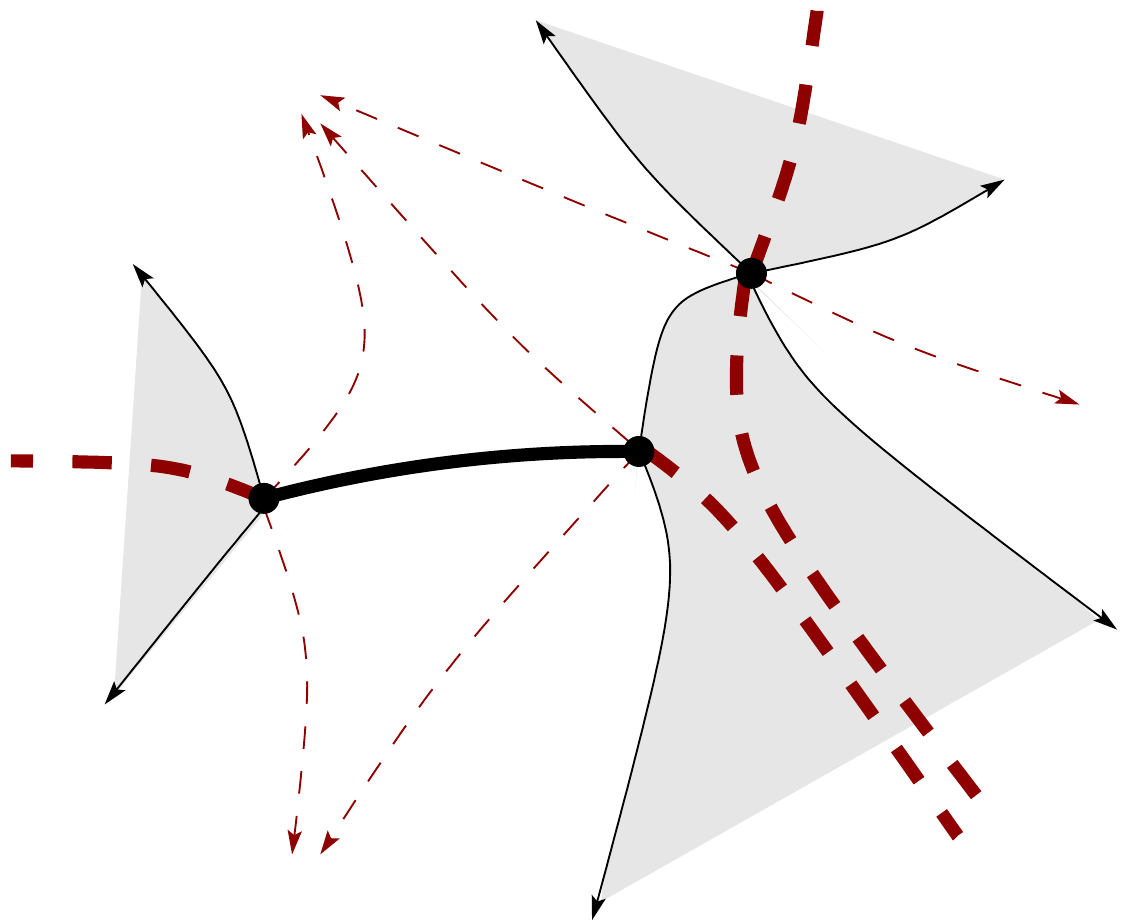}
		\caption{}
	\end{subfigure}
	\begin{subfigure}[b]{0.3 \textwidth}\centering
		{\includegraphics[scale=.21]{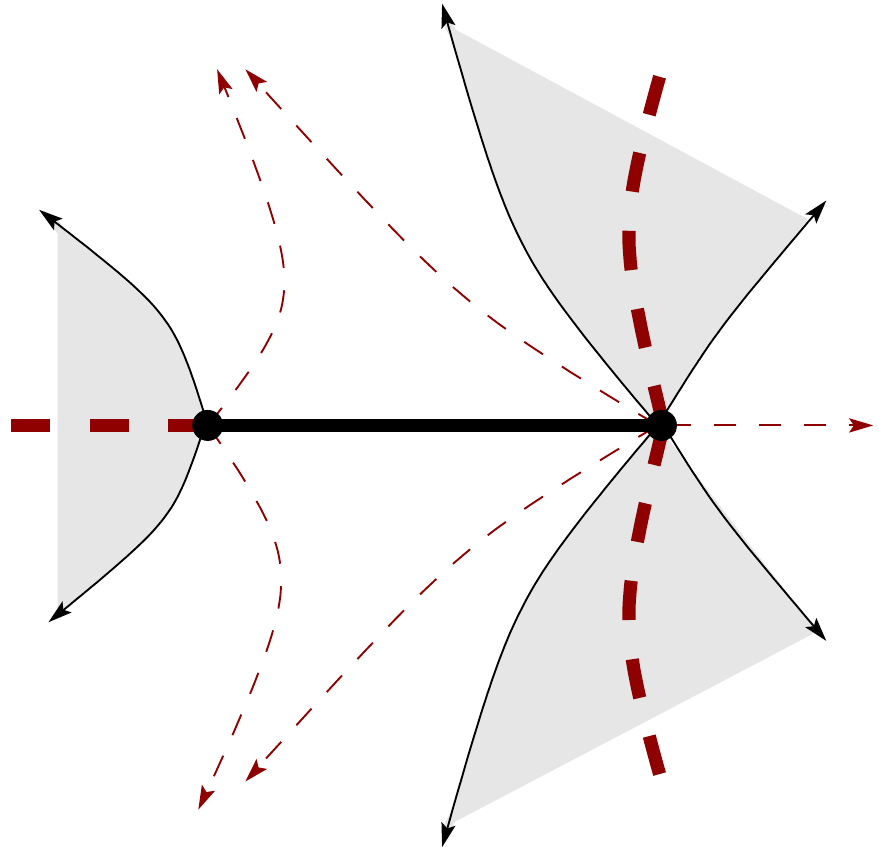}} 
		\caption{}
	\end{subfigure}
	\begin{subfigure}[b]{0.3 \textwidth}\centering
		{\includegraphics[scale=.19]{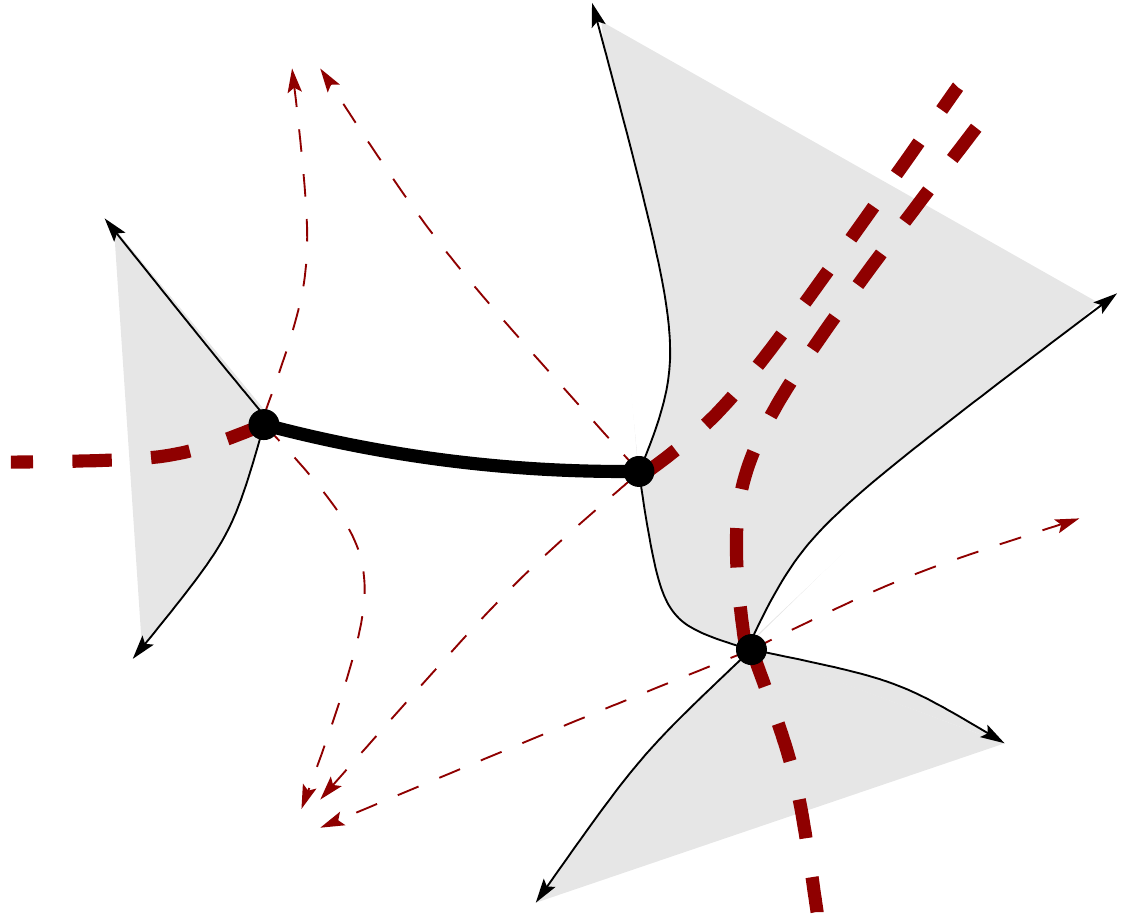}}
		\caption{}
	\end{subfigure}
	\caption{\small Schematic representation of the critical (solid) and critical orthogonal (dashed) graphs of $\varpi_t(z)$ when (a) \( t\in \partial O_{0,-\ii} \), $\im(t)>0$; (b) \( t=3 \cdot 2^{-2/3}\); (c) \( t\in \partial O_{0,-\ii} \), \( \im(t)<0 \). The bold curves represent the preferred S-curve $\Gamma_t$. Shaded region is the set $\{\mathcal U(z;t)<0\}$.}
	\label{s-curves2}
\end{figure} 

\begin{remark}
Theorem~\ref{thm:geometry1} below gives an ``appropriate'' choice for a chain of integration in \eqref{ortho}. Each provided chain can be easily modified within the region \( \{z:\mathcal U(z;t)<0\} \) so that its point set is connected and corresponds to an S-curve guaranteed by Theorem~\ref{fundamental}. That is, our choice modifies the S-curve only within \( \{z:\mathcal U(z;t)<0\} \), which is possible, see Remark~\ref{remark:freedom}, and allows us to choose the chain of integration that consists solely of trajectories and orthogonal trajectories of \( \varpi_t(z) \).
\end{remark}

\begin{figure}[ht!]
	\includegraphics[scale = 0.25]{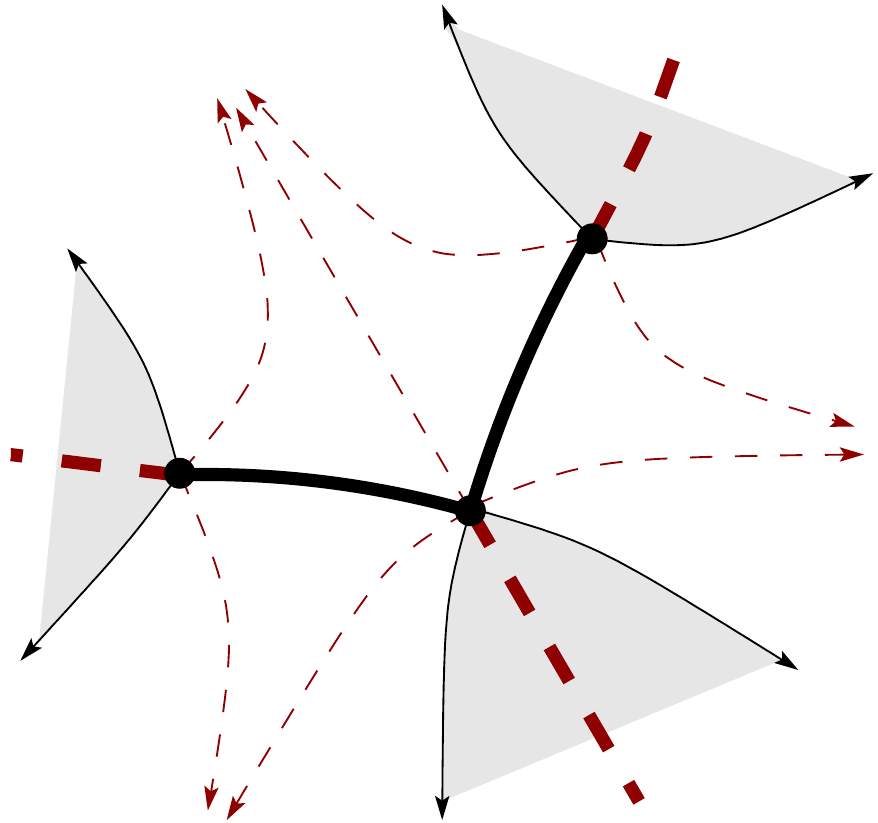}
	\caption{\small Schematic representation of the critical (solid) and critical orthogonal (dashed) graphs of $\varpi_t(z)$ when \( t\in (\partial O_{1,-\ii}\cap \partial O_{1,-})\setminus\{t_\mathsf{cr},\eta t_\mathsf{cr}\}  \). The bold curves represent the preferred S-curve $\Gamma_t$. Shaded region is the set $\{\mathcal U(z;t)<0\}$.}
	\label{s-curves4}
\end{figure}

\begin{remark}
Recall further  that to describe S-curves for \( t\notin O_1 \), it is enough to describe them for \( t\in \overline O_{0,-\ii} \) and \( t\in (\partial O_{1,-\ii}\cap \partial O_{1,-})\setminus\{t_\mathsf{cr},\eta t_\mathsf{cr}\} \), see Figure~\ref{fig:t-plane}, due to Remark~\ref{remark:symmetries}.
\end{remark}

\begin{theorem}
\label{thm:geometry1}
Let $\mu_t$ and $Q(z;t)$ be as in Theorem~\ref{fundamental}, $J_t=\mathrm{supp}(\mu_t)$. Assume that \( \alpha_0\alpha_1\alpha_2 \neq0 \). When $t\in\overline O_{0,-\ii}$, the polynomial $Q(z;t)$ is of the form \eqref{eq:Q-factorized-1}
with $a(t)$, $b(t)$, and $c(t)$  given by
\begin{equation}
\begin{cases}
a(t) &:= x_{-\ii}(t)-\ii\sqrt2/\sqrt{x_{-\ii}}(t), \smallskip \\
b(t) &:= x_{-\ii}(t)+\ii\sqrt2/\sqrt{x_{-\ii}}(t), \smallskip \\
c(t) &:= -x_{-\ii}(t),
\end{cases}
\label{eq:abc-t}
\end{equation}
where $\sqrt {x_{-\ii}}(t)$  is the branch holomorphic in $O_{(-\ii)}$ satisfying $\sqrt{x_{-\ii}}(0)= \ee^{\pi\ii/3}$. In these cases $J_t=\Gamma[a,b]$ and the chain of integration \( \Gamma \) in \eqref{ortho} can be chosen as
\[
\alpha_0 \left(\Gamma\big( \ee^{\pi\ii}\infty,a\big)\cup J_t \cup \Gamma\big(b, \ee^{-\pi\ii/3}\infty\big) \right) - \alpha_2 \left( \Gamma\big( \ee^{-\pi\ii/3}\infty,c\big]\cup \Gamma\big(c, \ee^{\pi\ii/3}\infty\big) \right)
\]
when $\im(t)>0$, see Figure~\ref{s-curves1}(A) and Figure~\ref{s-curves2}(A); as
\[
\alpha_0 \left(\Gamma\big( \ee^{\pi\ii}\infty,a\big) \cup J_t\cup \Gamma(b,c] \right) + \alpha_1 \Gamma\big( \ee^{-\pi\ii/3}\infty,c\big) - \alpha_2 \Gamma\big(c,\ee^{\pi\ii/3}\infty\big)
\]
when $t$ is real, see Figure~\ref{s-curves1}(B) and Figure~\ref{s-curves2}(B); 
\[
\alpha_0 \left(\Gamma\big( \ee^{\pi\ii}\infty,a\big) \cup J_t \cup \Gamma\big(b, \ee^{\pi\ii/3}\infty\big) \right) + \alpha_1 \left( \Gamma\big( \ee^{-\pi\ii/3}\infty,c\big]\cup \Gamma\big(c, \ee^{\pi\ii/3}\infty\big) \right)
\]
when $\im(t)<0$, see Figure~\ref{s-curves1}(C) and Figure~\ref{s-curves2}(C). 

If \( t\in (\partial O_{1,-\ii}\cap O_{1,-})\setminus\{t_\mathsf{cr},\eta t_\mathsf{cr}\} \), the support \( J_t=\Gamma[a,c] \cup \Gamma[c,b] \) is still a single Jordan arc and  the chain of integration \( \Gamma \)  in \eqref{ortho} can be chosen as
\[
\alpha_0 \left(\Gamma\big(\ee^{\pi\ii}\infty,a\big]\cup \Gamma(a,c)\right) + \alpha_1 \Gamma\big(\ee^{-\pi\ii/3}\infty,c\big] -\alpha_2 \left( \Gamma(c,b) \cup \Gamma\big[b,\ee^{\pi\ii/3}\infty\big) \right),
\]
 see Figure~\ref{s-curves4}.
\end{theorem}

These claims are part of \cite[Theorems~3.2 and 3.4]{MR3607591}, where a particular case $\lambda = -\ii$ (\( \alpha_0 = -\alpha_2 = 1/\pi \ii \), \( \alpha_1 = 0 \)) was analyzed. However, as one can clearly see, the obtained chains are homologous to \( \Gamma \) from \eqref{eq:Gamma}--\eqref{alphas}. In fact, Theorem~\ref{thm:geometry1} remains valid as stated for \( \lambda=-\ii \) as well, but  Remark~\ref{remark:symmetries} is not applicable to this case. So, S-curves for \( \lambda=-\ii \) and \( t\in O_{(-\ii)}\setminus O_{1,-\ii} \) are not obtained by rotation of the ones described above, see~\cite[Figure~4(a-c)]{MR3607591}.

\subsection{Two-cut Cases}

While it is possible to analytically continue $a(t), b(t)$, and $c(t)$ in \eqref{eq:abc-t} past the boundary of $O_{0,-\ii}$, the critical graph of $\varpi_t(z)$ as defined by \eqref{eq:Q-factorized-1} no longer admits a curve $\Gamma_t \in \mathcal T$ satisfying $\Gamma_t \subset \{z: \mathcal U(z;t) \leq 0\}$, see \cite[Theorems~3.2 and~3.4, and Figure 4]{MR3607591}. 
\begin{figure}[t]
	\includegraphics[scale = 0.25]{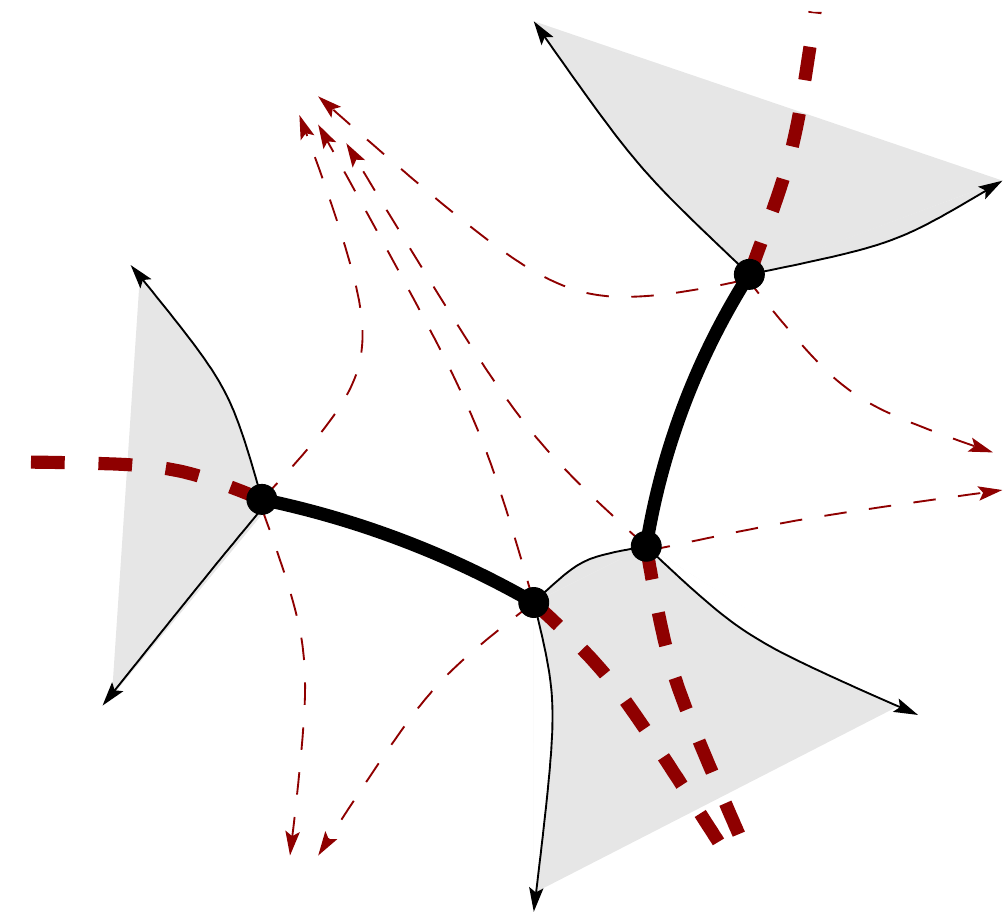}
	\caption{\small Schematic representation of the critical (solid) and critical orthogonal (dashed) graphs of $\varpi_t(z)$ when \( t\in O_{1,-\ii}  \). The bold curves represent the preferred S-curve $\Gamma_t$. Shaded region is the set $\{\mathcal U(z;t)<0\}$.}
	\label{s-curves5}
\end{figure}
Hence, we are forced to abandon ansatz \eqref{eq:Q-factorized-1} and have to look for \( Q(z;t) \) with four simple zeros. Conditions  \eqref{eq:abc-system} are now complemented by the \emph{Boutroux conditions}:
\begin{equation}
	\re \left(\int_{z_i(t)}^{z_j(t)} Q^{1/2}(s;t) \ \dd s \right)  = 0, \quad i,j \in \{0,1,2,3\},
	\label{eq:boutroux}
\end{equation}
for any continuous determination of \( Q^{1/2}(z;t) \) on the path of integration, where \( z_i(t) \) are the zeros of \( Q(z;t) \). Relations \eqref{eq:boutroux} follow from \eqref{eq:eq-measure}, which tells us that the integrals are purely imaginary on the branch cuts of \( Q^{1/2}(z;t) \), and \eqref{em2}, \eqref{em0}, which tell us that the integral between the cuts (if the cuts are not connected to each other) also must be purely imaginary so that the real constant \( \ell_{\Gamma_t} \) is the same on all (both) cuts. Notice that both remarks made before Theorem~\ref{thm:geometry1} remain relevant to the theorem below.

 \begin{theorem}
	\label{thm:geometry2}
	Let $\mu_t$ and $Q(z;t)$ be as in Theorem~\ref{fundamental}, $J_t=\mathrm{supp}(\mu_t)$. Assume that \( \alpha_0\alpha_1\alpha_2 \neq0 \). When $t\in O_{1,-\ii}$, the polynomial $Q(z;t)$ is of the form
	\[
	Q(z;t)=\frac14(z-a_1(t))(z-b_1(t))(z-a_2(t))(z-b_2(t))
	\]
	with $a_1(t)$, $b_1(t)$, \( a_2(t) \), and $b_2(t)$ all distinct. The real and imaginary parts of \( a_i(t),b_i(t) \) are real analytic functions of \( \re(t) \) and \( \im(t) \), however, at no point of \( O_{1,-\ii} \) any of the functions  \( a_i(t),b_i(t) \) is complex analytic. The chain of integration \( \Gamma \) in \eqref{ortho} can be chosen as
	\begin{multline*}
		\alpha_0\left(\Gamma\big(\ee^{\pi\ii}\infty,a_1\big) \cup \Gamma[a_1,b_1] \cup \Gamma\big(b_1,\ee^{-\pi\ii/3}\infty\big) \right) - \\ \alpha_2 \left( \Gamma\big(\ee^{-\pi\ii/3}\infty,a_2\big) \cup \Gamma[a_2,b_2] \cup \Gamma\big(b_2,\ee^{\pi\ii/3}\infty\big) \right), 
	\end{multline*}
see Figure~\ref{s-curves5} (this fixes our labeling of the zeros of \( Q(z;t) \)). Moreover, it holds that
	\begin{equation}
		\label{ts8a}
		a_1(t),b_1(t)\to c(t^*), \quad b_1(t),a_2(t) \to c(t^*), \qandq a_2(t),b_2(t) \to c(t^*)
	\end{equation}
	as \( t\to t^*\in \partial O_{1,-\ii} \) with \( t^*\in\partial O_{1,-} \), \( t^*\in \partial O_{1,-\ii} \), and \( t^*\in \partial O_{0,0} \), respectively.
\end{theorem}

This is  \cite[Theorem~3.2]{MR4436195}, where the case $\lambda = -\ii$  was analyzed. As before,  one can readily see that the obtained chains are homologous to \( \Gamma \) from \eqref{eq:Gamma}--\eqref{alphas}.

\subsection{Trefoil Cases}

It only remains to describe what happens when \( t\in O_{1,-} \). In this case the geometries of the critical and critical orthogonal graphs of \( \varpi_t(z) \) are distinct from all the previous cases.
\begin{figure}[t]
\includegraphics[scale=.2]{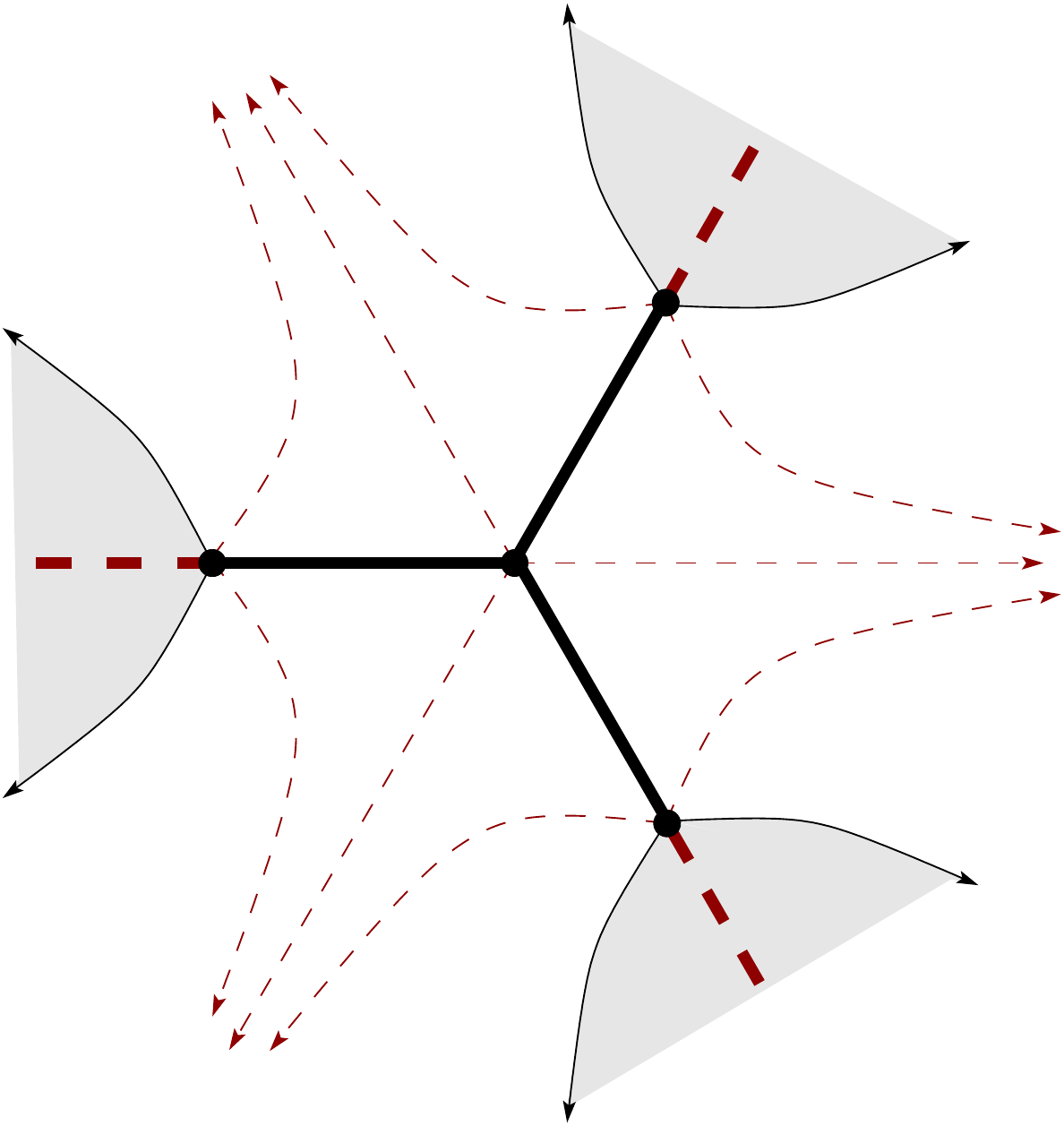}
\caption{\small Schematic representation of the critical (solid) and critical orthogonal (dashed) graphs of $\varpi_t(z)$ when \( t\in O_{1,-} \). The bold curves represent the preferred S-curve $\Gamma_t$. Shaded region is the set $\{\mathcal U(z;t)<0\}$.}
	\label{s-curves3}
\end{figure} 
Our approach essentially consists in observing that \( L_0\cup L_1\cup L_2 \) is the desired S-contour when \( t=0 \in O_{1,-} \) and arguing that the geometric structure of the critical and critical orthogonal graphs cannot change when \( t \) is varied within \( O_{1,-} \).

 \begin{theorem}
\label{thm:geometry3}
Let $\mu_t$ and $Q(z;t)$ be as in Theorem~\ref{fundamental}, $J_t=\mathrm{supp}(\mu_t)$. Assume that \( \alpha_0\alpha_1\alpha_2 \neq0 \). When $t\in O_{1,-}$, the polynomial $Q(z;t)$ is of the form
\[
Q(z;t)=\frac14(z-z_0(t))(z-z_1(t))(z-z_2(t))(z-z_3(t))
\]
with $z_i(t)$, \( i\in\{0,1,2,3\} \), all distinct. The real and imaginary parts of \( z_i(t) \) are real analytic functions of \( \re(t) \) and \( \im(t) \). The  chain of integration \( \Gamma \) in \eqref{ortho} can be chosen as
	\begin{multline*}
		\alpha_0\left(\Gamma\big(\ee^{\pi\ii}\infty,z_0\big) \cup \Gamma[z_0,z_3] \right) + \alpha_1\left(\Gamma\big(\ee^{-\pi\ii/3}\infty,z_1\big) \cup \Gamma[z_1,z_3) \right) + \\ \alpha_2 \left( \Gamma\big(\ee^{\pi\ii/3}\infty,z_2\big) \cup \Gamma[z_2,z_3) \right), 
	\end{multline*}
see Figure~\ref{s-curves3} (this fixes our labeling of the zeros of \( Q(z;t) \)). Moreover, it holds that
\[
z_0(t) \to a(t^*), \quad z_2(t)\to b(t^*), \qandq z_1(t),z_3(t) \to c(t^*)
\]
as \( t\to t^*\in \partial O_{1,-\ii}\cap \partial O_{1,-} \). 
\end{theorem}

The rest of this section is dedicated to the proof of Theorem~\ref{thm:geometry3}. Let us emphasize again that in the studied cases all the zeros of \( Q(z;t) \) must be distinct as coincidence of any two of them leads to the ansatz \eqref{eq:Q-factorized-1}, which necessarily implies that \( a(t),b(t),c(t) \) are as in \eqref{eq:abc-t}, where \( x_{-\ii}(t) \) can be replaced by \( x_0(t) \) or \( x_{\ii}(t) \), but no combination of the critical and critical orthogonal trajectories of the corresponding differentials form a desirable curve \( \Gamma_t \) (all the possible configurations were already examined before, see \cite[Theorems~3.2 and~3.4, and Figure~4]{MR3607591}).

\subsubsection{Step 1}

Since the Jacobian of the elementary symmetric functions of the zeros of \( Q(z;t) \) is its discriminant and all the zeros are distinct when \( t\in O_{1,-} \), it is enough to examine dependence on \( t \) of the coefficients of \( Q(z;t) \). In view of \eqref{eq:Q-unfactored}, we only need to study the free coefficient \( K(t) \).

Fix \( t^* \in O_{1,-} \) and let \( K^*=K(t^*) \) be the corresponding constant guaranteed by Theorem~\ref{fundamental} and Remark~\ref{remark:uniqueness}. Consider
\[
Q(z;t,K) := \dfrac{1}{4}(z^2 - t)^2 + z + K
\]
for \( t, K \) in some neighborhoods of \( t^* \) and \( K^* \), respectively. We take these neighborhoods small enough so that there are disjoint neighborhoods of \( z_i(t^*) \), \( i\in\{0,1,2,3 \} \), the zeros of \( Q(z;t^*) \), each containing one zero of \( Q(z;t,K) \). We label \( z_i(t,K) \) the zero of \( Q(z;t,K) \) that approaches \( z_i(t^*) \) as \( (t,K)\to(t^*,K^*) \) (these zeros are smooth functions of \( t \) and \( K \) as mentioned before). Define
\[
\RS_{t,K} := \left \{(z, w) \ : \ w^2 = Q(z;t,K) \right\} \subset \C^2.
\]
Since the zeros of \( Q(z;t,K) \) are simple, this is an elliptic, i.e., genus 1, Riemann surface. Select a homology basis on \( \RS_{t,K} \), say $(\boldsymbol \alpha_{t,K}, \boldsymbol \beta_{t,K})$. We can do it in such a way that the natural projection \( (z,w) \mapsto z \) of the cycle \( \boldsymbol\alpha_{t,K} \) is a Jordan arc connecting \( z_0(t,K) \) to \( z_1(t,K) \) while the natural projection of the cycle \( \boldsymbol\beta_{t,K} \) is a Jordan arc connecting \( z_0(t,K) \) to \( z_2(t,K) \). Moreover, we can arrange so that these Jordan arcs coincide for all considered \( (t,K) \) outside of small neighborhoods of \( z_0(t^*) \), \( z_1(t^*) \), and \( z_2(t^*) \). Set
\[
B_{\boldsymbol\gamma}(t,K) := \frac12\re\left(\oint_{\boldsymbol\gamma_{t,K}}w\dd z\right), \quad \boldsymbol\gamma\in\{\boldsymbol\alpha,\boldsymbol\beta\},
\]
which we consider to be functions of \( \re(t) \), \( \im(t) \), \( u:=\re(K) \), and \( v:=\im(K) \). We further make sure that the homology bases are selected so that \( B_{\boldsymbol\gamma}(t,K) \) are continuous functions of their parameters around \( (t^*,K^*) \). Boutroux conditions \eqref{eq:boutroux} yield that
\begin{equation}
	B_{\boldsymbol \alpha}(t,K) = 0 \qandq B_{\boldsymbol \beta}(t,K) = 0.
	\label{eq:boutroux-2}
\end{equation}
The Cauchy-Riemann equations  allow us to compute the Jacobian
\begin{align*}
	\det \dfrac{\partial (B_{\boldsymbol \alpha},B_{\boldsymbol \beta}) }{\partial (u, v)} &= \frac{1}{4} \left(- \re \left(   \oint_{\boldsymbol \alpha} \dfrac{\dd z}{w}   \right) \im  \left( \oint_{ \boldsymbol \beta} \dfrac{\dd z}{w}  \right) + \re  \left( \oint_{ \boldsymbol \beta} \dfrac{\dd z}{w}  \right) \im  \left( \oint_{\boldsymbol \alpha} \dfrac{\dd z}{w}  \right) \right)\\
	&= \frac{1}{4} \im \left( \oint_{ \boldsymbol \alpha} \dfrac{\dd z}{w} \overline{\oint_{ \boldsymbol \beta} \dfrac{\dd z}{w}} \right).
\end{align*}
The differential $\dd z /w$ is the unique (up to multiplication by a constant) holomorphic differential on $\RS_{t,K}$ and by Riemann's bilinear relation, see \cite[Lemma~35.2]{Bliss}, we have 
\[
\det \dfrac{\partial (B_{\boldsymbol \alpha},B_{\boldsymbol \beta}) }{\partial (u, v)} < 0.
\]
Thus, it follows from the Implicit Function Theorem that in some neighborhood of \( t^* \) there exists a function \( K^*(t) \) which is real analytic in \( \re(t) \) and \( \im(t) \) and such that the equations \eqref{eq:boutroux-2} hold with \( (t,K^*(t)) \) in this neighborhood of \( t^* \). We still need to argue that \( K(t) = K^*(t) \), To this end, we shall examine the critical graph of \( \varpi_t^*(z) := Q(z;t,K^*(t))\dd z^2\). Before we do this, let us recall relevant facts of the theory of quadratic differentials.

\subsubsection{Quadratic Differentials}

Given a quadratic differential \( \varpi_t^*(z) \) as above, it holds that
\begin{itemize}
	\item two distinct (orthogonal) trajectories meet only at critical points, see \cite[Theorem 5.5]{Strebel}, which, in this case,  are the zeros of \( Q(z;t,K^*(t)) \) and the point at infinity;
	\item no finite union of (orthogonal) trajectories can form a closed Jordan curve because $Q(z;t,K^*(t))$ is a polynomial, see \cite[Lemma 8.3]{Pomme};
	\item trajectories of $\varpi_t^*(z)$ cannot be recurrent (dense in two-dimensional regions), see \cite[Theorem 3.6]{Jenkins};
	\item since infinity is a pole of $\varpi_t^*(z)$ order 8, the critical trajectories can approach infinity only in six distinguished directions, namely, asymptotically to the lines \( \{\arg(z) = -5\pi/6+k\pi/3\} \), \( k\in\{0,1,2,3,4,5\} \);
\item there exists a neighborhood of infinity such that any trajectory entering it necessarily tends to infinity, see \cite[Theorem 7.4]{Strebel}.
\end{itemize}  

Denote by \( \mathcal G_t^* \) the critical graph of \( \varpi_t^*(z) \), that is, the totality of all the critical trajectories of \( \varpi_t^*(z) \). In the case of polynomial quadratic differentials, more can be said about the topological nature of $\C \setminus \mathcal G_t^*$. The complement of \( \mathcal G_t^* \) can be written as a disjoint union of either half-plane or strip domains, see \cite[Theorem~3.5]{Jenkins}. Recall that a \emph{half-plane (or end) domain} is swept by trajectories unbounded in both directions that approach infinity along consecutive critical directions. Its boundary is connected and consists of a union of two unbounded critical trajectories and a finite number (possibly zero) of short trajectories of $\varpi_t^*(z)$. The map $z \mapsto \int^z \sqrt{Q(s;t,K^*(t))} \dd s$ maps end domains conformally onto half planes $\{z \in \C \ | \ \re (z) >c \}$ for some $c \in \R$ that depends on the domain, and extends continuously to the boundary. Similarly, a \emph{strip domain} is swept by trajectories unbounded in both directions, but its boundary consists of two disjoint trajectories of $\varpi_t^*(z)$, each of which is comprised of two unbounded critical trajectories and a finite number (possibly zero) of short trajectories. The map $z \mapsto \int^z \sqrt{Q(s;t,K^*(t))} \dd s$ maps strip domains conformally onto vertical strips $\{w \in \C \ | \ c_1 < \re (w) < c_2 \}$ for some $c_1,c_2 \in \R$ depending on the domain, and extends continuously to their boundaries. The number $c_2 - c_1$ is known as the \emph{width of a strip domain} and can be calculated in terms of $\varpi_t^*(z)$ as 
\begin{equation}
	\label{strip-width}
	\left| \re \int_p^q \sqrt{Q(s;t,K^*(t))} \dd s \right|
\end{equation}
for any two points $p, q$ belonging to different components of the boundary of the domain.

\subsubsection{Step 2}

Equations \eqref{eq:boutroux-2} imply existence of three short critical trajectories of $ \varpi_t^*(z) $. Indeed, let \( J \) be a Jordan arc that connects a pair of zeros of $ Q(z;t,K^*(t)) $ and \( \boldsymbol\gamma \) be a cycle on \( \RS_{t,K^*(t)} \) whose natural projection is \( J \). Then it holds that
\[
\re\left(\int_JQ^{1/2}(s;t,K^*(t))\dd s\right) = \pm\frac12\re\left(\oint_{\boldsymbol\gamma }w \dd s\right),
\]
where \( Q^{1/2}(s;t,K^*(t)) \) is a continuous branch on \( J \) and the sign \( + \) or \( - \) depends on the choice of this branch as well as the chosen orientations for \( J \) and \( \boldsymbol\gamma \). It is well known, see \cite[Theorem~34.3]{Bliss}, that
\[
\oint_{\boldsymbol\gamma }w \dd s = n_1 \oint_{\boldsymbol\alpha_{t,K^*(t)} }w \dd s + n_2 \oint_{\boldsymbol\beta_{t,K^*(t)} }w \dd s  + 2\pi\ii \sum m_k  \,\mathrm{res}_{z_k} \,  (w\dd z),
\]
where \( n_1,n_2 \) and \( m_k \) are integers and the sum is taken over all the residues of the differential \( w\dd z \). Since the only poles of \( w\dd z \) are located at points on top of infinity with residues \( \pm 1 \), the last two relations together with \eqref{eq:boutroux-2} imply that
\begin{equation}
\label{width}
\re\left(\int_JQ^{1/2}(s;t,K^*(t))\dd s\right) = 0
\end{equation}
for any Jordan arc \( J \) connecting any pair of zeros of \( Q(s;t,K^*(t)) \) (that is, this polynomial has property \eqref{eq:boutroux}). Recall now that each zero $ z_i^*(t)=z_i(t,K^*(t)) $ of $ Q(z;t,K^*(t)) $ is incident with three critical trajectories of \( \varpi_t^*(z) \). If all three critical trajectories out of $ z_i^*(t)$ approach infinity, then $ z_i^*(t) $ must belong to a boundary of at least one strip domain. Let $ z_j^*(t) $ be a different zero belonging to the other component of the boundary of this strip domain. Then it follows from \eqref{strip-width} and \eqref{width} that the width of this strip domain is  $ 0 $, which is impossible. Thus, each zero  of $ Q(z;t,K^*(t)) $ must be coincident with at least one short trajectory. Therefore, either there is a zero connected by short trajectories to the remaining three zeros or there are at least two short trajectories connecting two pairs of zeros. In the latter case, label these zeros by $ z_0^*(t),z_1^*(t) $ and $ z_2^*(t),z_3^*(t) $. If the other two trajectories out of both $ z_0^*(t) $ and $ z_1^*(t) $ approach infinity, one of these zeros again must belong to the boundary of a strip domain with either $ z_2^*(t) $ or $ z_3^*(t) $ belonging to the other component of the boundary. As before, \eqref{strip-width} and \eqref{width} yield that the width of this strip domain is $ 0 $, which, again, is impossible. Thus, in this case there also exists a third short critical trajectory.

Since short critical trajectories cannot form a loop, they form a connected set, say \( \mathcal G_t^\mathrm{sh} \), that contains all four zeros of \(  Q(z;t,K^*(t)) \) (this set is either a trefoil or a single Jordan arc). Define
\begin{equation}
\label{eq:U_star}
\mathcal U^*(z;t) := \re\left(2\int_e^z Q^{1/2}(s;t,K^*(t)) \dd s \right),
\end{equation}
where \( e\in \mathcal G_t^\mathrm{sh} \) and \( Q^{1/2}(s;t,K^*(t)) \) is the branch holomorphic in \( \C\setminus \mathcal G_t^\mathrm{sh} \) that behaves like \( \frac12z^2 + O(z) \)  as \( z \to \infty \). \( \mathcal U^*(z;t) \) is a well defined harmonic function in \( \C\setminus \mathcal G_t^\mathrm{sh} \) by \eqref{width} (which is independent of the choice of \( e \)). Moreover, this function is either harmonic across a given short critical trajectory in \( \mathcal G_t^\mathrm{sh} \) (this happens to the ``middle section'' when three short critical trajectories form a single Jordan arc) or can be harmonically continued across it by \( - \mathcal U^*(z;t) \). Since the zeros of \( Q^{1/2}(s;t,K^*(t)) \) converge to the zeros of \( Q^{1/2}(s;t^*) \) as \( t\to t^* \), we have that \( \mathcal U^*(z;t) \) converges to \( \mathcal U(z;t^*) \) point-wise. Using the above harmonic continuation to define a single harmonic function on a double ramified cover of this disk via concatenation of \( \mathcal U^*(z;t) \) and \( - \mathcal U^*(z;t) \) and applying the maximum modulus principle and \cite[Theorem 1.3.10]{MR1334766}, we deduce that the functions \( \mathcal U^*(z;t) \) converge to \( \mathcal U(z;t^*) \) as \( t\to t^* \) uniformly on compact subsets of $\C$. Furthermore, we have that
\begin{multline}
\label{eq:at_infty}
 \mathcal U^*(z;t) = -\re(V(z;t)) + 2\log|z/e| + \\ \re\left(2\int_e^z \left (Q^{1/2}(s;t,K^*(t)) -\frac{s^2-t}2 - \frac1s \right) \dd s \right),
\end{multline}
where the integrand behaves as \( O(z^{-2}) \) at infinity and therefore the last summand above is harmonic at infinity. Thus, using \eqref{eq:at_infty} we have that the functions \( \mathcal U^*(z;t) \) converge to \( \mathcal U(z;t^*) \) as \( t\to t^* \) uniformly on the whole sphere \( \overline\C \). In particular, since the critical graph of \( \varpi_t^*(z) \) is the zero level set of \( \mathcal U^*(z;t) \), this function can be either positive and negative on different sides of a trajectory (i.e., it is harmonic across) or be positive on both sides (i.e., it is subharmonic across) since this is true for \( \mathcal U(z;t^*) \) by \eqref{em2} and \eqref{em0}. The latter situation happens on the part of \( \mathcal G_t^\mathrm{sh} \) that serves as the branch cut for  \( Q^{1/2}(s;t,K^*(t)) \), which we denote by \( J^*_t \).

\subsubsection{Step 3}

Now, we would like to identify possible structures of the critical graph of $\varpi_t^*(z)$. To this end, we shall use what is known as \emph{clock diagrams}.  
\begin{figure}[t]
	\begin{subfigure}[b]{0.22 \textwidth}\centering
		{\includegraphics[scale=.17]{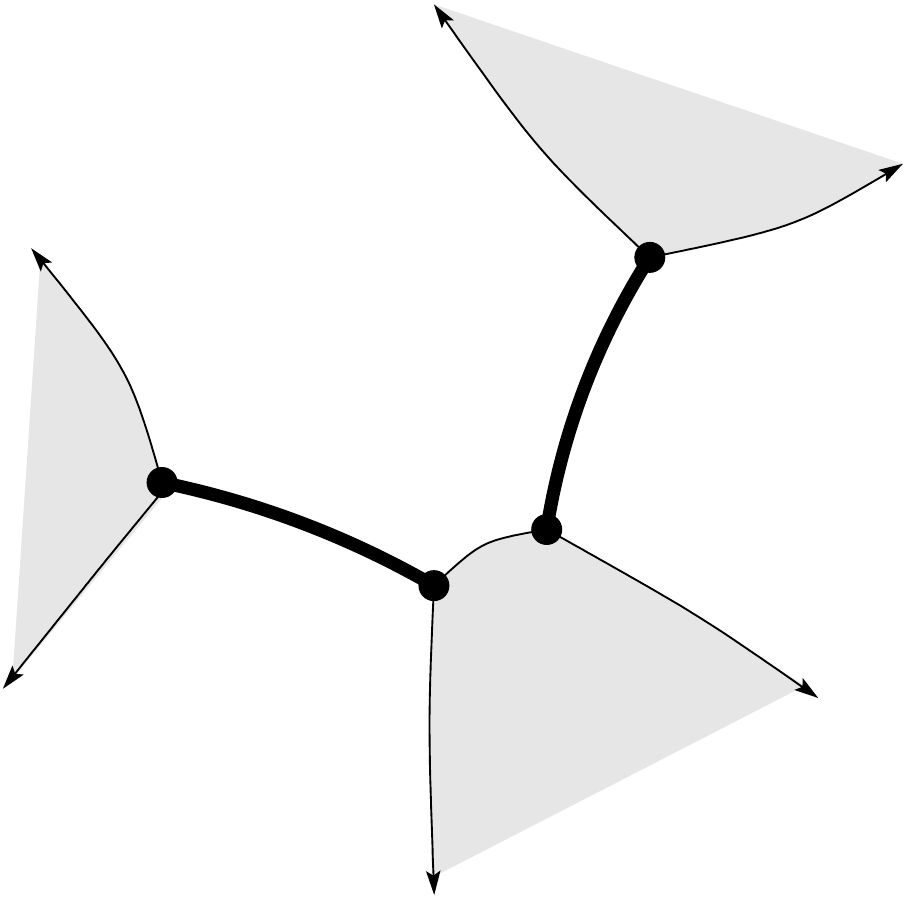}} 
		\caption{}
	\end{subfigure}
	\begin{subfigure}[b]{0.22 \textwidth}\centering
		\includegraphics[scale=.25]{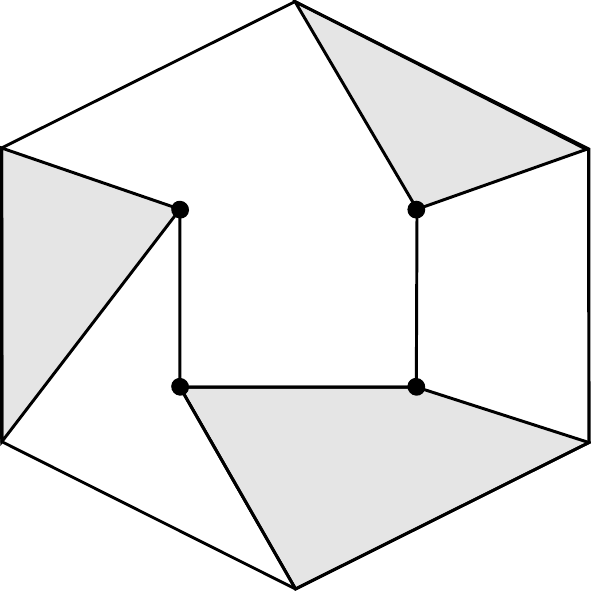}
		\caption{}
	\end{subfigure}
		\begin{subfigure}[b]{0.22 \textwidth}\centering
		{\includegraphics[scale=.1]{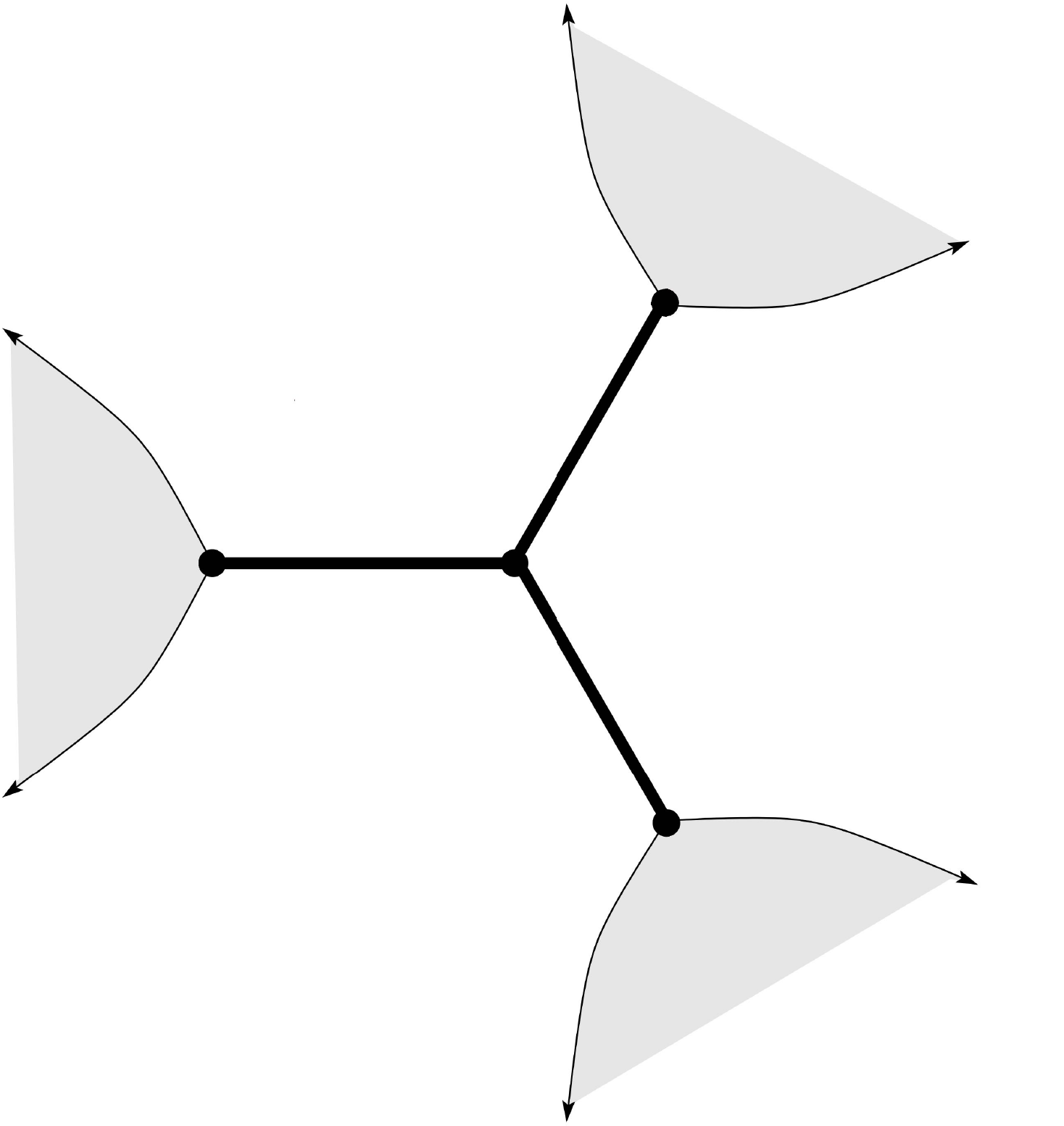}} 
		\caption{}
	\end{subfigure}
	\begin{subfigure}[b]{0.22 \textwidth}\centering
		\includegraphics[scale=.25]{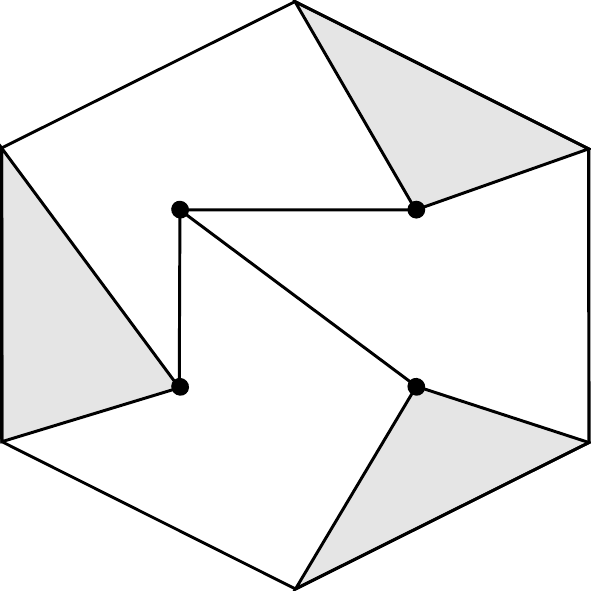}
		\caption{}
	\end{subfigure}
	\caption{\small Schematic representation of $\mathcal G_t^*$ (A) or (C) and its associated clock diagram (B) or (D). Shaded region is the set $\{\mathcal U(z;t)<0\}$.}
	\label{fig:clock}
\end{figure} 
Introduced in \cite[Definition~5.4]{BM09}, (topological) clock diagrams are a schematic way of representing the critical graph of $\varpi_t^*(z)$ that allows for a more combinatorial treatment similar in spirit to \cite{B11} though more concrete since we deal with a fixed polynomial potential. In our setting, a clock diagram consists of the outer hexagon whose vertices correspond to the distinguished directions at infinity; inner vertices that correspond to the finite critical points, i.e., zeros of \( Q(z;t,K^*(t)) \), that are connected to each other and the vertices of the hexagon by edges corresponding to the critical  trajectories of $\varpi_t^*(z)$; and finally the shaded regions that correspond to the strip or end domains where \( \mathcal U^*(z,t)< 0 \). It follows from \eqref{em0} that every other edge of the hexagon must border a shaded region starting with the vertical edge on the left, which corresponds to the sector between the distinguished direction \( 5\pi/6 \) and \( 7\pi/6 \), see Figure~\ref{fig:clock}.

\begin{figure}[t]
	\begin{subfigure}[b]{0.24 \textwidth}\centering
		\includegraphics[scale = 0.3]{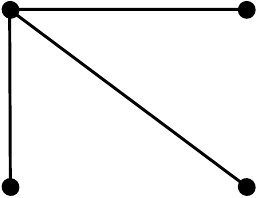}
		\caption{}
	\end{subfigure}
	\begin{subfigure}[b]{0.24 \textwidth}\centering
		\includegraphics[scale = 0.3]{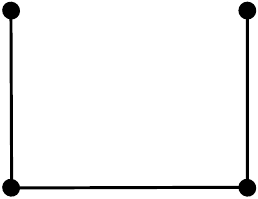}
		\caption{}
	\end{subfigure}
	\begin{subfigure}[b]{0.24 \textwidth}\centering
		\includegraphics[scale = 0.3]{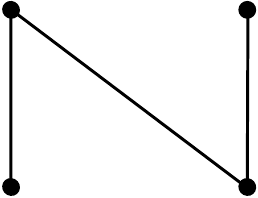}
	\caption{}
	\end{subfigure}
	\begin{subfigure}[b]{0.24 \textwidth}\centering
		\includegraphics[scale = 0.3]{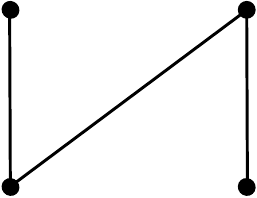}
	\caption{}
	\end{subfigure}
	\caption{\small The four possible configurations of $\mathcal G_t^\mathrm{sh}$ up to rotation.}
	\label{fig:finite-config}
\end{figure}

\begin{figure}[t]
	\begin{subfigure}[b]{0.15 \textwidth}\centering
		\includegraphics[scale = 0.175]{good-clock-2.pdf}
		\caption{}
	\end{subfigure}
	\begin{subfigure}[b]{0.15 \textwidth}\centering
		\includegraphics[scale = 0.175]{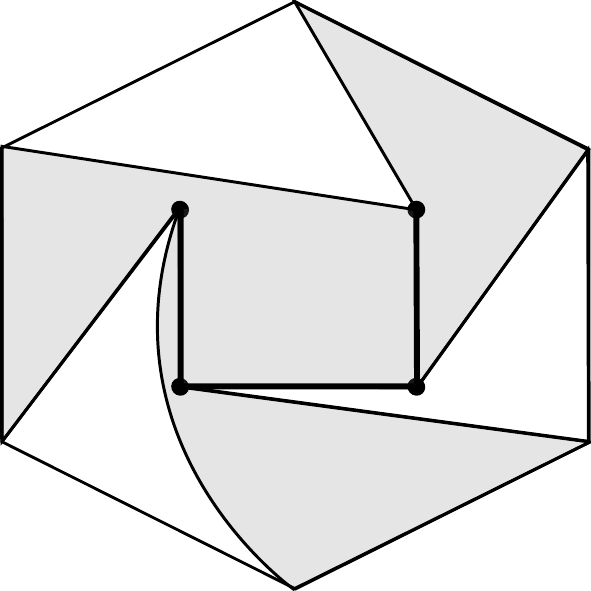}
		\caption{}
	\end{subfigure}
	\begin{subfigure}[b]{0.15 \textwidth}\centering
		\includegraphics[scale = 0.175]{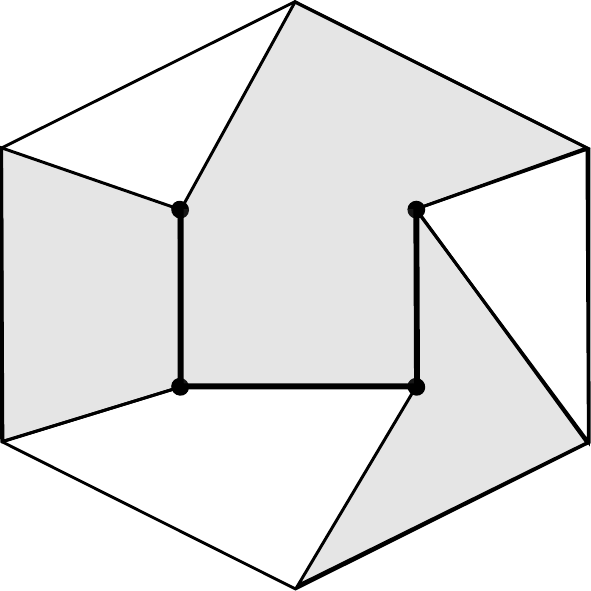}
	\caption{}
	\end{subfigure}
	\begin{subfigure}[b]{0.15 \textwidth}\centering
		\includegraphics[scale = 0.175]{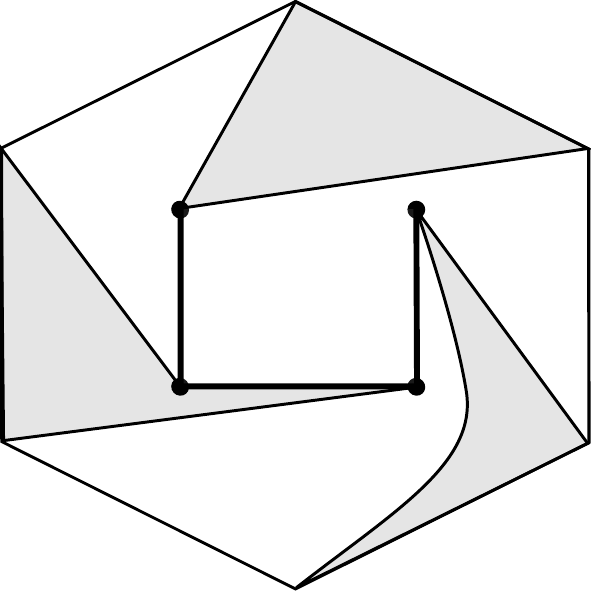}
	\caption{}
	\end{subfigure}
		\begin{subfigure}[b]{0.15 \textwidth}\centering
		\includegraphics[scale = 0.175]{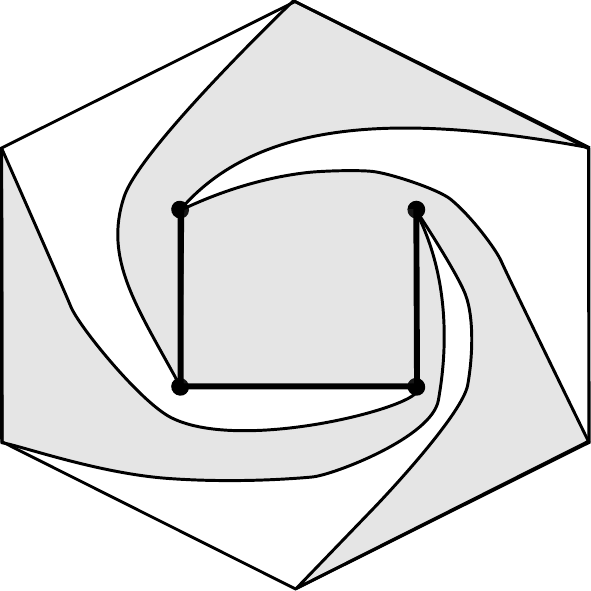}
	\caption{}
	\end{subfigure}
	\begin{subfigure}[b]{0.15 \textwidth}\centering
		\includegraphics[scale = 0.175]{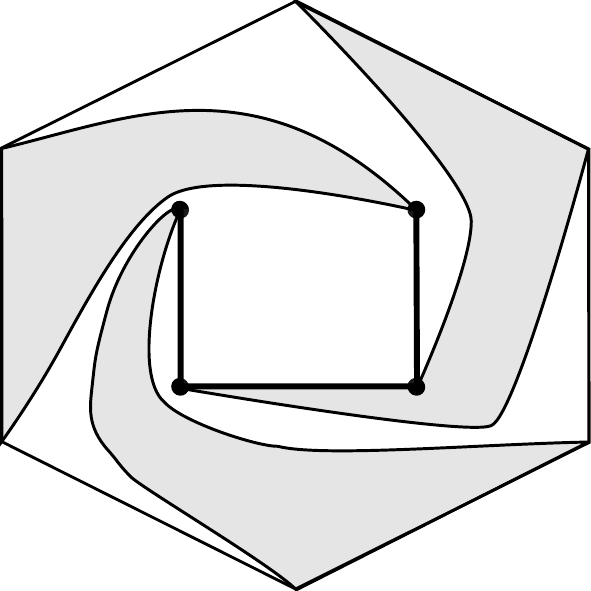}
	\caption{}
	\end{subfigure}
	\caption{\small The six possible ways to connect $\mathcal G_t^\mathrm{sh}$ from Figure~\ref{fig:finite-config}(B) to the hexagon.}
	\label{fig:6ways}
\end{figure}

\begin{figure}[t]
	\begin{subfigure}[b]{0.15 \textwidth}\centering
		{\includegraphics[scale=.175]{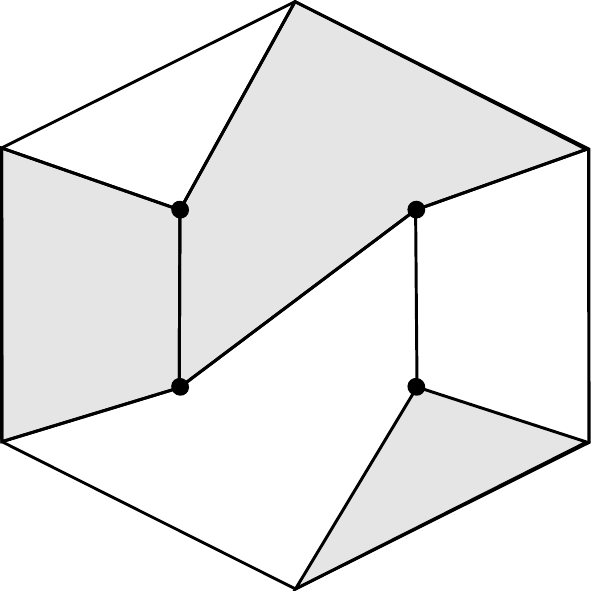}} 
		\caption{}
	\end{subfigure}
	\begin{subfigure}[b]{0.15 \textwidth}\centering
		\includegraphics[scale=.175]{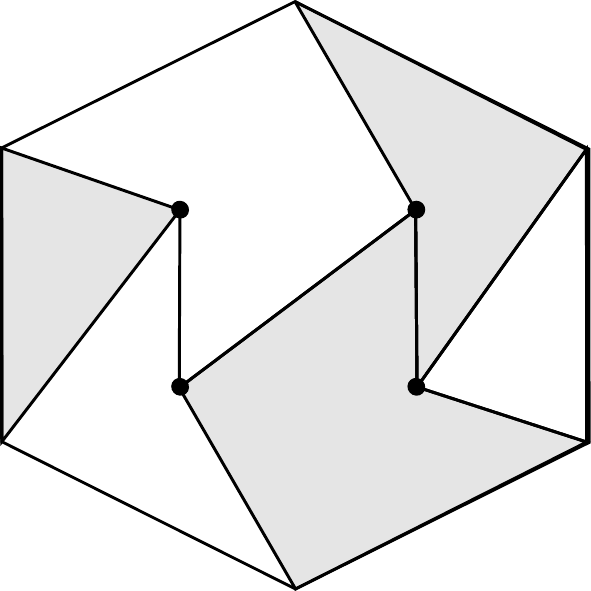}
		\caption{}
	\end{subfigure}
	\begin{subfigure}[b]{0.15 \textwidth}\centering
		{\includegraphics[scale=.175]{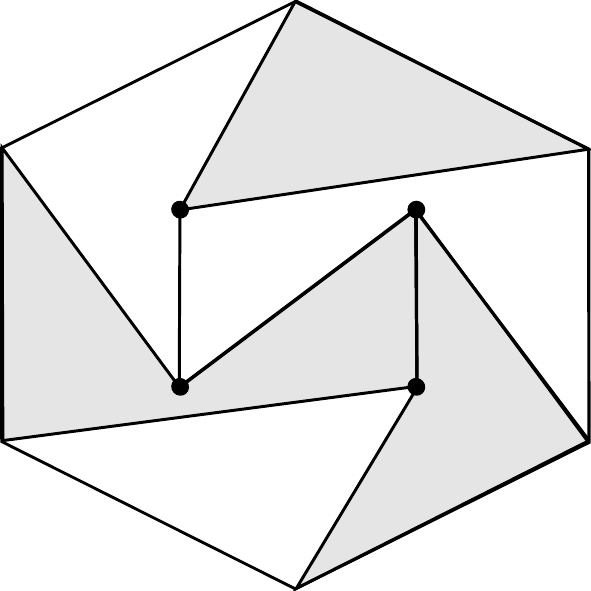}} 
		\caption{}
	\end{subfigure}
	\begin{subfigure}[b]{0.15 \textwidth}\centering
		\includegraphics[scale=.175]{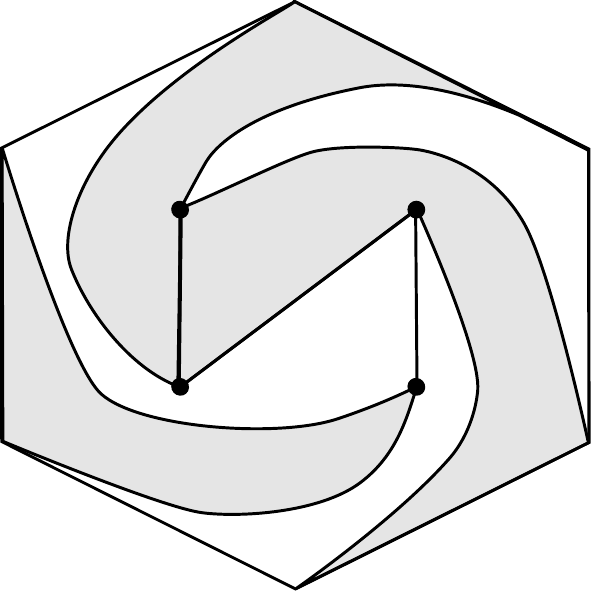}
		\caption{}
	\end{subfigure}
	\begin{subfigure}[b]{0.15 \textwidth}\centering
		{\includegraphics[scale=.175]{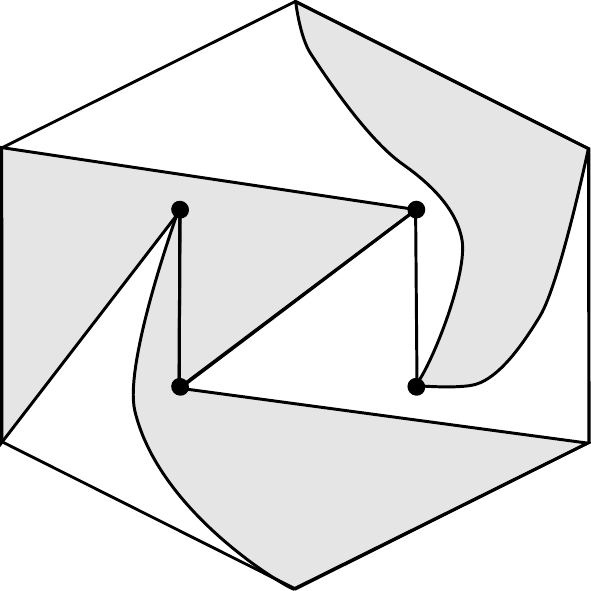}} 
		\caption{}
	\end{subfigure}
	\begin{subfigure}[b]{0.15 \textwidth}\centering
		\includegraphics[scale=.175]{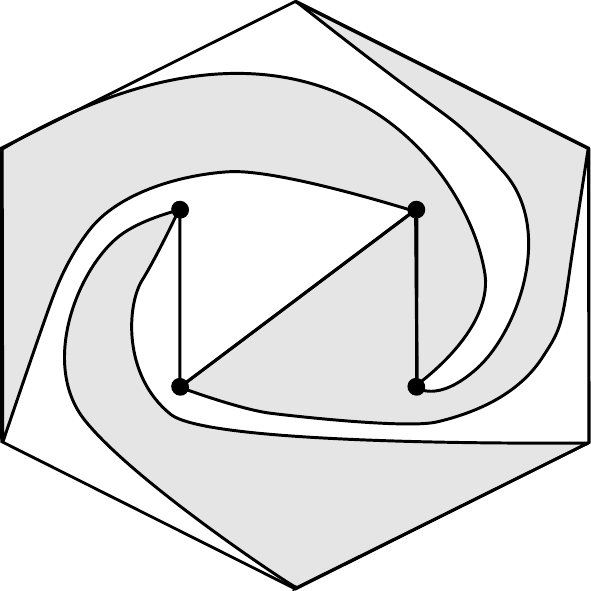}
		\caption{}
	\end{subfigure}		
	\caption{\small The six possible ways to connect $\mathcal G_t^\mathrm{sh}$ from Figure~\ref{fig:finite-config}(D) to the hexagon.}
	\label{fig:bad-clock}
\end{figure}

Since the critical graph of \( \varpi_t^*(z) \) must contain exactly three short trajectories not allowing for loops, we immediately see that there are exactly 4 possible configurations of short trajectories (up to rotation), shown on Figure \ref{fig:finite-config}. Each of these configurations can be attached to the boundary of the hexagon in 6 different ways since once the direction in which one trajectory goes to infinity is fixed, the others are uniquely determined by Teichm\"uller's Lemma, see \cite[Theorem 14.1]{Strebel}, see Figures~\ref{fig:6ways} and~\ref{fig:bad-clock}. 
However, we now observe the importance of the shading. As explained after \eqref{eq:at_infty}, $J_t^*$ must border unshaded regions on both sides whereas any other trajectory must be the boundary of both a shaded and an unshaded region.  Hence, of the 6 possible ways to connect the finite critical points to the hexagon, only 3 allow for a valid coloring when $\mathcal G_t^\mathrm{sh}$ is as on Figures~\ref{fig:finite-config}(A,B) (and their rotations), see Figure~\ref{fig:6ways}, while
there are no valid colorings when $\mathcal G_t^\mathrm{sh}$ is as on Figures~\ref{fig:finite-config}(C,D), see Figure~\ref{fig:bad-clock}. In the sequel, we will need to talk about clock diagrams and critical graphs which are ``structurally the same." To do so in our setting, we attach to a critical graph $\mathcal{G}$ with critical points $z_1, z_2, z_3, z_4$ the adjacency data
\[
\left \{ \{z_1, A_1\}, \{z_2, A_2 \}, \{z_3, A_3 \}, \{z_4, A_4 \} \right \},
\]
where $A_i \subset \N$ is a finite set and $k \in A_i$ if and only if $z_i$ is on the boundary of the end domain containing the line $ \{ r\ee^{\pi k\ii/3} \ : \ r > R\}$ for $R$ large enough.   
	
\begin{definition}
	We say that two critical graphs are {\it structurally equivalent} if there exists a labeling of the critical points such that both graphs have the same adjacency data. Two clock diagrams are structurally equivalent if their corresponding critical graphs are structurally equivalent.
\end{definition}

\begin{figure}[t]
	\begin{subfigure}[b]{0.24 \textwidth}\centering
		\includegraphics[scale=.2]{good-clock-2.pdf}
		\caption{}
	\end{subfigure}
	\begin{subfigure}[b]{0.24 \textwidth}\centering
		{\includegraphics[scale=.2]{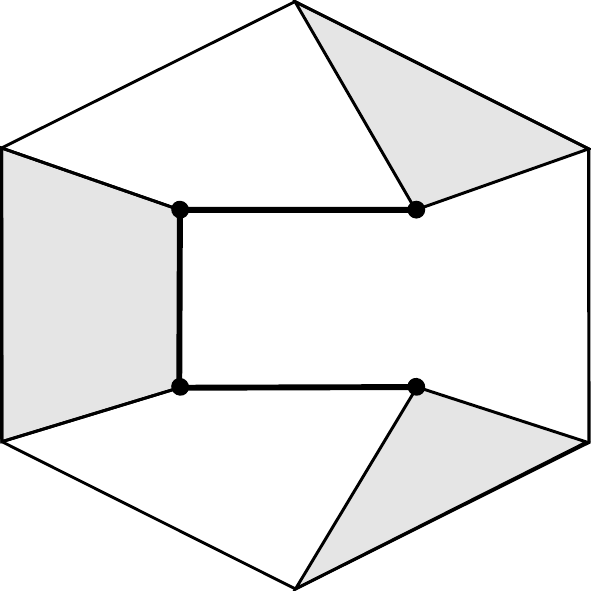}} 
		\caption{}
	\end{subfigure}
	\begin{subfigure}[b]{0.24 \textwidth}\centering
		\includegraphics[scale=.2]{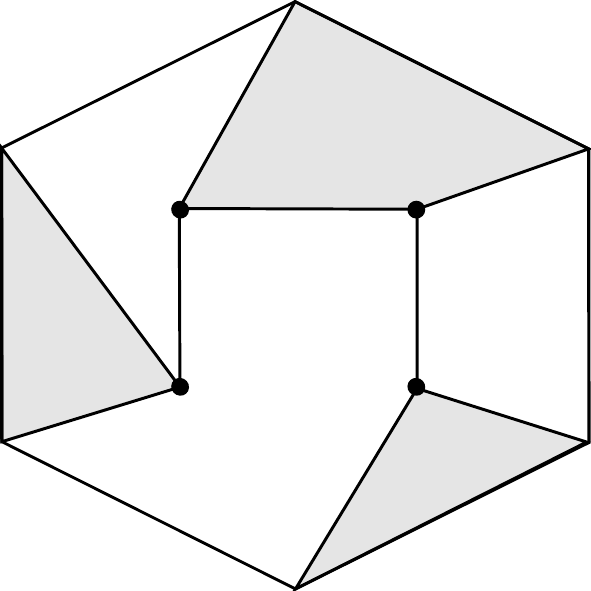}
		\caption{}
	\end{subfigure}
\begin{subfigure}[b]{0.24 \textwidth}\centering
	\includegraphics[scale=.2]{good-clock-4.pdf}
	\caption{}
\end{subfigure}
	\caption{\small The only four admissible clock diagrams. }
	\label{fig:good-clock}
\end{figure}

With this definition in mind, observe that the clock diagrams on Figures~\ref{fig:6ways}(A,D,F) structurally describe the same critical graphs as diagrams on Figures~\ref{fig:good-clock}(A,B,C), respectively. Hence, if we were to draw the admissible diagrams corresponding to a rotation of \( \mathcal G_t^\mathrm{sh} \) as on Figure~\ref{fig:finite-config}(B), they would again structurally correspond to the critical graphs depicted on Figures~\ref{fig:good-clock}(A,B,C). As all admissible diagrams corresponding to \( \mathcal G_t^\mathrm{sh} \) as on Figure~\ref{fig:finite-config}(A) are structurally the same, see Figure~\ref{fig:good-clock}(D), Figure~\ref{fig:good-clock} depicts all possible admissible clock diagrams for the critical graph of \( \varpi_t^*(z) \).

\subsubsection{Step 4}

Since \( \mathcal U^*(z;t) \) is subharmonic in \( \C \) and harmonic in \( \C\setminus J_t^* \), its Laplacian is a positive measure, say \( 2\mu_t^* \), supported on \( J_t^* \). It follows from \eqref{eq:at_infty}, the Riesz Decomposition Theorem \cite[Theorem~3.7.9]{MR1334766} and Liouville's Theorem for subharmonic functions \cite[Corollary~2.3.4]{MR1334766} to identify the harmonic part that
\[
\mathcal U^*(z;t) = -\re(V(z;t)) + \ell_t^* - 2U^{\mu_t^*}(z)
\]
for some real constant \( \ell_t^* \). The measure \( \mu_t^* \) has mass \( 1 \) as this is the only case when \( 2U^{\mu_t^*}(z) \) behaves like \( -2\log|z| \) as \( z\to\infty \), see \eqref{eq:at_infty}. Moreover, as we have mentioned before, the very definition of \( \mathcal U^*(z;t) \) in  \eqref{eq:U_star} implies that \( -\mathcal U^*(z;t) \) is the harmonic continuation of \( \mathcal U^*(z;t) \) across any subarc of \( J_t^* \). This, however, is equivalent to saying that \( \mathcal U^*(z;t) \) satisfies \eqref{em4}. Hence, if we add to the corresponding \( \mathcal G_t^\mathrm{sh} \), of which \( J_t^* \) is a part, three unbounded Jordan arcs that lie within \( \{\mathcal U^*(z;t)<0\} \) (shaded regions on Figure~\ref{fig:good-clock}) to create an element of \( \mathcal T \), we will obtain a symmetric contour. Recall now that the uniqueness in Remark~\ref{remark:uniqueness} was not derived based on the maximization of the minimal energy, which is the defining property of \( J_t \) in Theorem~\ref{fundamental}, but based on being an S-curve. Hence, \( J_t^* \) is \( J_t \) and \( \mu_t^* \) and \( \mu_t \). Moreover, this finishes the proof of real analytic dependence of the zeros of \( Q(z;t) \) on \( \re(t) \) and \( \im(t) \).

Since the function \( \mathcal U(z;t) \) converge uniformly to \( \mathcal U(z;t^*) \) as \( t\to t^* \) and the critical graph of \( \varpi_t(z) \) can have only one of the four structural forms depicted on Figure~\ref{fig:good-clock}, the critical graphs of \( \varpi_t(z) \) are structurally the same as the critical graph of \( \varpi_{t^*}(z) \) for all \( t \) in some small neighborhood of \( t^* \). As any two points in \( O_{1,-} \) can be joined by a path that can be covered by finitely many such neighborhoods, the structure of the critical graph of \( \varpi_t(z) \) must be the same for every \( t\in O_{1,-} \). It readily follows from Remark~\ref{remark:symmetries} that \( K(0)=0 \) and therefore
\[
\varpi_0(z) = \frac{z(z^3+4)}4 \dd z^2,
\]
whose critical graph can be readily determined and is depicted on Figure~\ref{s-curves3}. Since this structure of the critical graph uniquely determines the structure of the critical orthogonal graph, this finishes the proof of the theorem except for the last claim.

\subsubsection{Step 5}

It follows from \cite[Theorem 5.11]{MR3589917} that if $ t $ remains in a bounded set, so do the zeros of $ Q(z;t) $. Fix $t^* \in \partial O_{1,-}$ and let $ \{t_m\} $ be a sequence such that $ t_m\to t^* $ as $ m\to\infty $. Restricting to a subsequence if necessary, we see that there exist $z_i^*$ such that $ z_i(t_m)\to z_i^* $ as $ m\to\infty $, \( i\in\{0,1,2,3\} \), where $ z_i(t) $ are the zeros of \( Q(z;t) \). Clearly, polynomials $ Q(z;t_m) $ converge uniformly on compact subsets of $ \C $ to $ Q^*(z) $, the polynomial with zeros $ z_i^* $ and leading coefficient $ 1/4 $. With the help of uniform convergence we can conclude that \eqref{width} must be satisfied with \( Q(z;t,K^*(t)) \) replaced by \( Q^*(z) \). At this point, we can repeat the steps that follow \eqref{width} (we will need to add clock diagrams corresponding to possible confluence of the zeros of \( Q^*(z) \)) to conclude that an S-curve can be constructed out of the critical and critical orthogonal trajectories of \( Q^*(z)\dd z^2 \), and then appeal to Remark~\ref{remark:uniqueness} to conclude that \( Q^*(z) = Q(z;t^*) \).

\bibliographystyle{plain}

\bibliography{airyII}

\end{document}